\documentclass[iop]{emulateapj}
\usepackage{natbib}
\usepackage{color}
\usepackage{amsmath}
\usepackage{graphicx}
\usepackage{romannum}
\usepackage{xfrac}
\usepackage{threeparttable}

\shorttitle{Environmental Influences on Galaxy Gas Reservoirs}
\shortauthors{Stark et al.}

\begin{document}

\pagenumbering{arabic}

\title{The RESOLVE Survey Atomic Gas Census and Environmental Influences on Galaxy Gas Reservoirs}
\author{David V. Stark\altaffilmark{1,2,$\dagger$}, Sheila J. Kannappan
\altaffilmark{1,7}, Kathleen D. Eckert\altaffilmark{1,8}, Jonathan Florez\altaffilmark{3,4}, Kirsten R. Hall\altaffilmark{1,5}, Linda C. Watson\altaffilmark{6}, Erik A. Hoversten\altaffilmark{1}, Joseph N. Burchett\altaffilmark{1,7}, David T. Guynn\altaffilmark{1}, Ashley D. Baker\altaffilmark{1,8}, Amanda J. Moffett\altaffilmark{1,9}, Andreas A. Berlind\altaffilmark{3},  Mark A. Norris \altaffilmark{14}, Martha P. Haynes\altaffilmark{10}, Riccardo Giovanelli\altaffilmark{10}, Adam K. Leroy\altaffilmark{11}, D. J. Pisano\altaffilmark{12,16},  Lisa H. Wei\altaffilmark{13}, Roberto E. Gonzalez\altaffilmark{15}, Victor F. Calderon\altaffilmark{3}}
\altaffiltext{$\dagger$}{david.stark@ipmu.jp}
\altaffiltext{1}{Physics and Astronomy Department, University of North Carolina, Chapel Hill, NC 27516, USA}
\altaffiltext{2}{Kavli Institute for the Physics and Mathematics of the Universe (WPI), UTIAS, The University of Tokyo, Kashiwa, Chiba 277-8583, Japan}
\altaffiltext{3}{Department of Physics and Astronomy, Vanderbilt University, Nashville, TN 37240 USA}
\altaffiltext{4}{Fisk University, Nashville, TN 37208, USA}
\altaffiltext{5}{Department of Physics and Astronomy, Johns Hopkins University, Baltimore, MD 21218, USA}
\altaffiltext{6}{European Southern Observatory, Santiago, Chile}
\altaffiltext{7}{Department of Astronomy, University of Massachusetts, Amherst, MA 01003, USA }
\altaffiltext{8}{Department of Physics and Astronomy, University of Pennsylvania, Philadelphia, PA 19104, USA}
\altaffiltext{9}{International Centre for Radio Astronomy Research, University of Western Australia Perth, WA 6009, Australia}
\altaffiltext{10}{Center for Radiophysics and Space Research, Cornell University, Ithaca, NY 14853, USA} 
\altaffiltext{11}{Department of Astronomy, the Ohio State University, Columbus, OH 43210, USA}
\altaffiltext{12}{Department of Physics and Astronomy, West Virginia University, Morgantown, WV 26506, USA}
\altaffiltext{13}{Atmospheric and Environmental Research, Lexington, MA 02421, USA}
\altaffiltext{14}{Jeremiah Horrocks Institute, University of Central Lancashire, Preston, Lancashire, PR1 2HE, United Kingdom}
\altaffiltext{15}{Instituto de Astrof{\'i}sica, Pontificia Universidad Cat{\'o}lica de Chile, Santiago, Chile}
\altaffiltext{16}{Center for Gravitational Waves and Cosmology, West Virginia University, Chestnut Ridge Research Building, Morgantown, WV 26505}


\newcommand\myion[2]{#1\,{\scriptsize\rmfamily\Romannum{#2}}\relax}%
\newcommand{\HI}{\myion{H}{1}}   
\newcommand{\htwo}{H$_{2}$}
\newcommand{\htwohi}{H$_{2}$/H\,{\scriptsize I}}
\newcommand{\hmpc}{{\rm h}^{-1}\,{\rm Mpc}}
\newcommand{\msun}{M_{\odot}}

\begin{abstract}
  We present the \HI{} mass inventory for the REsolved Spectroscopy of a Local VolumE (RESOLVE) survey, a volume-limited, multi-wavelength census of $>$1500 $z=0$ galaxies spanning diverse environments and complete in baryonic mass down to dwarfs of $\sim$10$^9\,\msun$. This first 21cm data release provides robust detections or strong upper limits (1.4$M_{\rm HI}$ $<$ 5--10$\%$ of stellar mass $M_*$) for $\sim$94\% of RESOLVE. We examine global atomic gas-to-stellar mass ratios ($G/S$) in relation to galaxy environment using several metrics: group dark matter halo mass $M_h$, central/satellite designation, relative mass density of the cosmic web, and distance to nearest massive group. We find that at fixed $M_*$, satellites have decreasing $G/S$ with increasing $M_h$ starting clearly at $M_h\sim10^{12}\,\msun$, suggesting the presence of starvation and/or stripping mechanisms associated with halo gas heating in intermediate-mass groups. \textcolor{black}{The analogous relationship for centrals is uncertain because halo abundance matching builds in relationships between central $G/S$, stellar mass, and halo mass, which depend on the integrated group property used as a proxy for halo mass (stellar or baryonic mass)}. On larger scales $G/S$ trends are less sensitive to the abundance matching method.  At fixed $M_h\le10^{12}\,\msun$, the fraction of gas-poor centrals increases with large-scale structure density. In overdense regions, we identify a rare population of gas-poor centrals in low-mass \mbox{($M_h<10^{11.4}\,\msun$)} halos primarily located within $\sim$1.5$\times$ the virial radius of more massive ($M_h>10^{12}\,\msun$) halos, \textcolor{black}{suggesting that gas stripping and/or starvation may be induced by interactions with larger halos or the surrounding cosmic web. We find that the detailed relationship between $G/S$ and environment varies when we examine different subvolumes of RESOLVE independently, which we suggest may be a signature of assembly bias.}
\end{abstract}


\section{Introduction}
\label{sec:ch4:intro}
Galaxy gas reservoirs are the raw fuel for star formation and thus play a key role in galaxy evolution. \textcolor{black}{Galaxies are not isolated, but are subject to interactions with both other galaxies and the intergalactic medium (IGM)}.  Therefore, understanding the extent to which environment governs galaxy gas content is a fundamental ingredient to understanding galaxy assembly as a whole.  Several studies have highlighted the link between star formation and environment through the color-density relation,  which translates into the physical understanding that galaxies in dense regions have lower star formation rates (SFRs) and typically older ages than those in the field \citep{Kennicutt83,Gomez03,Cooper10}.  Likewise, galaxies in dense environments show gas deficiencies \citep{Davies73,Haynes84b,Solanes01,Cortese11,Catinella13} while the most gas-rich galaxies are often found in the least dense environments  \citep{Meyer07,Martin12}.

There are multiple possible connections between galaxy gas supply and the surrounding environment. For example, the low cold gas content among galaxies in dense environments can be attributed to mechanisms that cut off gas replenishment (i.e., starvation; \citealt{Larson80, Balogh00, Bekki02,Kawata08,Hearin16}) or directly remove gas (e.g., tidal, ram-pressure, or viscous stripping; \citealt{Gunn72}, \citealt{Nulsen82}, \citealt{Kenney04}). In the absence of these processes, galaxies acquire gas from their surroundings over time.  Fresh gas infall is needed to explain the roughly constant star formation history of the Milky Way \citep{Twarog80}, as well as the heavy element abundances in its stellar populations \citep{Chiappini01}.  Regular (and possibly overwhelming) gas infall  also explains the high gas content and exponential stellar mass growth of many dwarf galaxies in the local universe \citep{Kannappan13}, and there are multiple examples of early-type galaxies that appear to be (re)growing gas and stellar disks \citep{Cortese09, KGB, Lemonias11, Moffett12, Salim12, Stark13}. 

While galaxies can acquire new gas through hierarchical merging \citep{Lacey93}, a more subtle but extremely important alternative mechanism is the smooth accretion of the IGM, i.e. ``cosmological accretion.'' Traditional theory suggests that as gas enters a dark matter halo, it shock heats to the halo's virial temperature before slowly cooling onto the galaxy \citep{Rees77,Silk77,White78}.  Below a halo mass threshold, the cooling timescale may be short enough that infalling gas can avoid shock heating to the virial temperature \citep{White91,Birnboim03,Keres05,Dekel06,Keres09}.  This ``cold mode" of accretion is thought to take the form of gas streams that penetrate into halos along cosmic filaments, depositing cool gas onto galaxies much more rapidly than the traditional ``hot" mode.  

Direct detection of cool gas streams associated with cold mode accretion is difficult since this gas is expected to be in a low-density, warm-hot ionized state that lacks detectable emission at low redshift \citep{Bregman07}. However, a number of high-redshift studies have detected gas through Lyman-$\alpha$ emission or absorption with properties consistent with cold-mode accretion \citep{Nilsson06,Ribaudo11,Kacprzak12,Bouche13,Crighton13,Martin15}, and some absorption features consistent with pristine gas infall have also been reported at low redshift (e.g., \citealt{Burchett13}). Further evidence comes from observations of neutral atomic hydrogen (\HI{}) emission around nearby galaxies.  High-velocity clouds have been observed around many galaxies in the Local Group, particularly the Milky Way, and some of these clouds may have external origins    \citep{Wakker97,Sembach03,Braun04}.  

Key group halo mass scales theoretically associated with changes in accretion can be related to observed trends in galaxy properties.  The halo mass scale below which cold-mode accretion is expected to dominate over hot-mode accretion 
(${\sim}10^{11.4}\,\msun$; \citealt{Keres09}) matches the observed ``gas-richness threshold scale" \citep{Kannappan13}, where gas-dominated galaxies become the norm.  The scale above which cold-mode accretion is no longer present (${\sim}10^{12}\,\msun$; \citealt{Keres09}) matches the ``bimodality mass," which marks a transition between star-forming and ``quenched" galaxies \citep{Kauffmann03,Kannappan13}.  More recent simulations suggest that cold-mode accretion may be less important than previously thought, with infalling streams likely getting disrupted in the inner halo before reaching the central galaxy \citep{Nelson13}.  However, this effect is at least somewhat balanced by a faster cooling rate for gas accreted via the hot mode.

Recent work has often emphasized a picture wherein galaxy gas reservoirs are largely governed by dark matter halos and their internal environments: gas accretion rates are expected to be closely tied to the masses of dark matter halos, as are many processes that deplete gas content (e.g., ram pressure stripping; \citealt{Hester06}).  However, there is evidence that galaxy properties may also depend on the environment beyond the halo virial radius. \textcolor{black}{\citet{Kauffmann13} find that galaxy star formation rates (SFR) can be correlated on scales up to $\sim$4 Mpc (particularly for low-mass, low-SFR galaxies), well beyond the typical virial radii of individual groups. \citet{Lietzen12} find that groups at fixed richness have more passive galaxies if they reside in supercluster environments  as opposed to less dense environments, and \citet{Wang13} find that passive, low-mass group centrals are more strongly clustered than star-forming centrals of similar mass. Several studies have also shown that very low-density/void environments have larger fractions of low-mass, gas-rich, high specific star formation rate (sSFR) galaxies compared to non-void environments, and when the luminosity distributions of void/non-void samples are matched, void galaxies show on average bluer colors and higher sSFRs \citep{Grogin99,Rojas04,Rojas05,VonBendaBeckmann08,Hoyle12,Moorman14,Moorman15,Moorman16,Jones16}. Both \citet{Kreckel12} and \citet{Moorman16} show hints that void galaxies may have higher star formation efficiencies (defined as \mbox{${\rm SFR}/M_{\rm HI}$)}, although these findings are not statistically significant, and \citet{Beygu16} find that star formation efficiencies in voids are generally consistent with those in higher-density environments.}

\textcolor{black}{Large-scale environmental trends may reflect the phenomenon known as ``assembly bias," i.e., the dependence of the spatial distribution of halos not only on mass, but also assembly history \citep{Gao05}. A key aspect of assembly bias is that halos in overdense regions formed earlier, which may influence the properties of their galaxies.  Galaxies in underdense regions, having formed later, may have more gas than those galaxies which formed earlier in high-density regions and have had their gas supplies cut off \citep{Grogin00, Rojas04, Rojas05}.} A number of different physical mechanisms have been proposed that either remove gas or slow the infall of gas in dark matter halos in overdense regions.  Such environments may have higher rates of flyby interactions (involving ``ejected satellites" or ``splashback galaxies"), wherein a galaxy enters a more massive halo, loses its gas content, and then escapes the inner regions, at least temporarily \citep{Hansen09,Sinha12,Lu12,Rasmussen12,Wetzel12,Wetzel14}. Additionally, \citet{Bahe13} suggest that the IGM in large-scale structure leads to ram pressure stripping of hot halo gas (particularly at $M_*<10^{10}\,M_{\odot}$), reducing the potential of galaxies to replenish their cold gas supply.
Halo gas accretion rates may also be lessened by competition between dark matter halos \citep{Hearin16}, or by longer cooling times caused by earlier heating from the gravitational collapse of cosmic structure \citep{Cen11} and/or early active galactic nucleus (AGN) feedback \citep{Kauffmann15}.


In this work, we present the first 21cm data release for the REsolved Spectroscopy of a Local VolumE (RESOLVE) survey, a new multi-wavelength volume-limited census of galaxies in the local universe that has a large dynamic range of group halo masses ($10^{11-14}\,\msun$) and large-scale structure densities (factor of ${\sim}1000$ variation), and probes galaxy masses down to the dwarf galaxy regime (baryonic mass ${\sim}10^{9}$).  RESOLVE and its \HI{} census are ideally suited for environmental studies of global \HI-to-stellar mass ratios enabling us to address multiple key questions relating to the physical processes governing galaxy fuel supplies: how does gas content scale with halo mass?  Does this scaling behave differently for centrals and satellites? Does the observed gas deficiency previously observed in large groups and clusters also occur in more moderately sized dark matter halos? \textcolor{black}{How do the large-scale environments beyond group dark matter halos regulate galaxy gas content?  }

In \S\ref{sec:ch4:data}, we describe the RESOLVE survey and its 21cm census, followed by a discussion of the metrics \textcolor{black}{used to parametrize group dark matter halos and their larger-scale environment (halo mass, cosmic web density, and distance to the nearest massive group).  In \S\ref{sec:ch4:results}, we explore the influence of group halo mass on the gas content of central and satellite galaxies, while also highlighting possible biases introduced when estimating halo masses using different abundance matching prescriptions.  We also investigate the influence of environment on scales larger than dark matter halos by examining the relationship between gas content and both the relative density of large-scale structure and the distance to the nearest massive group, while also discussing how our results are affected by cosmic variance.} In \S\ref{sec:ch4:discussion} we interpret our findings from the point of view of the physical processes occurring within and around group dark matter halos and large-scale structure.  We summarize our conclusions in \S\ref{sec:ch4:conclusion}.

\section{Data and Methods}
\label{sec:ch4:data}

\subsection{The RESOLVE Survey}
\label{sec:ch4:resolve}

The RESOLVE survey\footnote{https://resolve.astro.unc.edu} is a volume-limited census of galaxies in the local universe with the goal of
accounting for baryonic and dark matter mass within a statistically complete subset of the $z=0$ galaxy population.  A complete description of the survey
design will be presented in S. J. Kannappan et al. (in prep), but we briefly
summarize the key aspects of the survey here.

\subsubsection{Survey Definition}
\label{sec:ch4:survey_def}

RESOLVE covers two equatorial strips, denoted ``RESOLVE-A" and ``RESOLVE-B," whose combined volumes total ${\sim}52,000\,{\rm Mpc}^3$. \mbox{RESOLVE-A} spans R.A. = 8.75h to 15.75h and decl. = 0$^{\circ}$ to 5$^{\circ}$, and \mbox{RESOLVE-B} spans from \mbox{R.A. = 22h to 3h} and \mbox{decl. = -1.25$^{\circ}$ to $+$1.25$^{\circ}$}. Both
regions are bounded in Local Group-corrected heliocentric velocity
from \mbox{$V_{LG}$=4500--7000 km s$^{-1}$}.  Final survey membership is based on the redshift of the group to which each galaxy is assigned (see \S\ref{sec:ch4:mhalo}) to avoid cases where peculiar velocities artificially push galaxies inside or outside the nominal RESOLVE volume. The RESOLVE survey benefits from a variety of multi-wavelength data.  This paper presents new 21cm observations, but an optical spectroscopic survey is under way, primarily
with the SOAR 4.1m telescope, and also using SALT, Gemini, and the AAT.
These observations provide either stellar or ionized gas kinematics in addition to gas
and stellar metallicities. RESOLVE also overlaps with several photometric surveys spanning near infrared to ultraviolet wavelengths, which are used to estimate colors and stellar masses (see \S \ref{sec:ch4:photometry} and \citealt{Eckert15}).

RESOLVE is designed to be baryonic mass limited as opposed to limited in stellar mass or luminosity. We define baryonic mass as $M_{\rm bary}=M_*+1.4M_{\rm HI}$, where $M_*$ is the stellar mass and $1.4M_{\rm HI}$ is the atomic hydrogen gas mass corrected for the contribution from helium.  We ignore the contribution from molecular hydrogen (H$_2$) in the cold gas budget.  H$_2$ may be a significant gas component for intermediate-mass spirals, but for our dwarf-dominated sample, we expect it to be negligible (see \citealt{Kannappan13}).  The baryonic mass is chosen to define the sample since it is a more fundamental characterization of {\it total} galaxy mass than is stellar mass, e.g., as seen in the necessity to include gas mass to obtain a linear Baryonic Tully-Fisher relation (BTFR) \citep{McGaugh00}, or the close association between the observed transitions in galaxy gas fractions and morphologies with {\it baryonic}, not stellar, mass scales \citep{Kannappan13}.  
 
The RESOLVE sample is initially selected on \mbox{$r$-band} absolute magnitude ($M_r$), since $r$-band magnitude closely correlates with total baryonic mass \citep{Kannappan13}.  By combining the SDSS redshift survey \citep{Abazajian09} with the Updated Zwicky Catalog (UZC; \citealt{Falco99}), HyperLEDA \citep{Paturel03}, 2dF \citep{Colless01}, 6dF \citep{Jones09}, GAMA \citep{Driver11}, Arecibo Legacy Fast ALFA (ALFALFA) \citep{Haynes11} \and new redshift observations with the SOAR and SALT telescopes (S. J. Kannappan et al. in prep), we obtain $r$-band completeness limits of $M_r<-17.33$ and $M_r<-17$ in RESOLVE-A and RESOLVE-B, respectively (the latter completeness limit being dimmer largely due to the overlap with the deep Stripe-82 SDSS field). The baryonic mass completeness limit is then estimated by considering the range of possible baryonic mass-to-light ratios at the $M_r$ completeness limit, which yields baryonic mass completeness limits of $M_{\rm bary}=10^{9.3}\,M_{\odot}$ and $M_{\rm bary}=10^{9.1}\,M_{\odot}$ in RESOLVE-A and RESOLVE-B, respectively \citep{Eckert16}.  Since gas mass information was not available for all galaxies at the start of the RESOLVE survey, indirect gas mass estimators (see \S\ref{sec:ch4:indirect_gs}, \citealt{Eckert15}) were used to identify objects with $r$-band magnitudes below the nominal completeness limit but with potentially high baryonic mass-to-light ratios.  Any such objects lacking gas information were targeted for 21cm follow-up to improve RESOLVE's baryonic mass completeness. 

Throughout this paper, we often use a stellar mass-limited sample since it tends to more clearly highlight processes that drive gas deficiency. The stellar mass completeness limits for RESOLVE are determined in the same fashion as the baryonic mass completeness limits, yielding limits of $M_*=10^{8.9}\,\msun$ and $M_*=10^{8.7}\,\msun$ in RESOLVE-A and RESOLVE-B, respectively.

\subsubsection{Custom Photometry and Stellar Masses}
\label{sec:ch4:photometry}

The photometric analysis for RESOLVE is fully described in \citet{Eckert15}. To briefly summarize, all photometric data, including SDSS {\it ugriz} \citep{Aihara11}, 2MASS {\it JHK} \citep{Skrutskie06}, UKIDSS {\it YHK} \citep{Hambly08}, {\it GALEX} NUV \citep{Morrissey07}, and Swift NUV \citep{Roming05}, have been reprocessed through custom pipelines to yield uniform magnitude measurements and improved recovery of low surface brightness emission (i.e., dwarf galaxies and outer disks).  Total magnitudes are calculated using multiple techniques to enable realistic uncertainty estimates.  

The new uniform photometry is used to calculate stellar masses (used extensively in this work) using the spectral energy distribution fitting code described in \citet{Kannappan07} and modified in \citet{Kannappan13}.  We use the second model grid from \citet{Kannappan13} which combines old simple stellar populations with age ranging from 2 to 12 Gyr and young stellar populations described either by continuous star formation from 1015 Myr ago until between 0 and 195 Myr ago, or by a simple stellar population with age 360, 509, 641, 806, or 1015 Myr.  For each model, the stellar mass is calculated and given a likelihood based on the $\chi^2$ of the model fit.  The stellar masses and likelihoods are then combined into a likelihood weighted stellar mass distribution, and the median of this distribution is used as the final stellar mass.  The stellar masses are given in \citet{Eckert15}.

\subsection{21cm Data}
\label{sec:ch4:21cm_data}

The goal of the RESOLVE 21cm census 
is to obtain strong detections (integrated S/N $>$ 5--10) or upper limits
(${\rm 1.4M_{\rm HI}/M_* < 0.05-0.1}$) for the atomic gas reservoirs of all galaxies in the sample. In the following sections, we describe the sources of our 21cm
data, resulting products, and the current status of the census.

\subsubsection{ALFALFA and Other Literature Data}
\label{sec:ch4:alfalfa}

The ALFALFA survey \citep{Giovanelli05}
overlaps 85\% of the RESOLVE footprint (only lacking coverage in
RESOLVE-B at \mbox{decl. $< 0^{\circ}$)}, and provides data satisfying our sensitivity requirements for
$\sim$65\% of the galaxies within this overlap region, or
$\sim$55\% of the entire RESOLVE survey.  The blindly detected 21cm sources in
the standard ALFALFA catalog are cross-matched with RESOLVE using a
match radius of 2\arcmin, corresponding to the spatial resolution of
the final ALFALFA data cubes.  Additionally, we search the ALFALFA data
cubes at the positions of all galaxies that lack counterparts within
the standard ALFALFA catalogs.  Their spectra are extracted using a
4\arcmin$\times$4\arcmin$\,$ box and provide upper limits (which are
not standard ALFALFA pipeline outputs) and in some cases, weak
detections.  The majority of the detections have signal-to-noise ratio S/N$<$5 and
some were found to be spurious, so most were followed up with single-dish
observations.

The other major source of literature data for RESOLVE comes from the large compilation of 21cm observations
presented in \citet{Springob05}.  We adopt their
fluxes corrected for beam offsets and source extent, but without the
corrections for \HI{} self-absorption, which are expected to  be no larger than 30\% for the most inclined systems \citep{Giovanelli94}.

\subsubsection{New Green Bank Telescope and Arecibo Observations }
\label{sec:ch4:gbt_ao}

To complete the RESOLVE \HI{} census, new 21cm observations were carried out with
the Robert C. Byrd Green Bank Telescope (GBT; programs 11B-056, 13A-276,
13B-246, 14A-441) and Arecibo Observatory (programs a2671, a2812, a2852).  GBT data were acquired over a total of 738 hr between August 2011 and July 2014. Observations were conducted in standard position switching
mode with typical scan lengths of five minutes.  We used the L-band receiver and the
GBT Spectrometer with a bandwidth of 50 MHz, spectral resolution of 1
kHz, and 9-level sampling (the VEGAS backend was briefly used while the GBT
Spectrometer was undergoing maintenance).  At the beginning of each
run, a bright quasar was observed to calibrate the data and check the
telescope pointing.

The close proximity of our targets provided opportunities to boost the
efficiency of our GBT observations. For galaxies within a few degrees
of each other and separated in heliocentric velocity by $>$1000 km
s$^{-1}$, a scan centered on one galaxy could serve as the OFF position for
a scan centered on the nearby galaxy (and vice versa), allowing us to
cut our total observing time for those targets in half.  We also
conducted observations where two galaxies shared the same OFF position
located midway between them, reducing total integration times by $\sim$30\%.  This observing strategy did not severely degrade the quality of our baselines.

Arecibo data were acquired over a total of \textcolor{black}{554} hr in March 2012 and again between July 2013 and May 2016.  Observations were done in standard position-switching mode
using scan lengths between three and five minutes.  We used the L-band
Wide receiver and the interim correlator with a bandwidth of \mbox{12.5
MHz}, \mbox{2 kHz} spectral resolution, and 9-level sampling.  Data were
calibrated by observing an internal noise diode of known temperature
before and after each scan.

\subsection{21cm Line Profile Analysis}
\label{sec:ch4:profile_analysis}

All new single-dish observations were reduced following standard GBT
and Arecibo pipeline IDL software packages.  Baselines, typically of
order 3--5, were fit to the emission free regions of each spectrum,
and the spectra were boxcar smoothed to a final velocity resolution of
$\sim$5.25 km s$^{-1}$.  For details on the reduction of the ALFALFA
and other literature data, we refer the reader to \citet{Haynes11} and
\citet{Springob05}.

\subsubsection{Atomic gas Mass}
\label{sec:ch4:gasmass}

Integrated 21cm line fluxes are measured by summing the channels
within the line profile.  The channels included in the integration
are judged by eye for each case.  The uncertainty on each flux
measurement is given by
\begin{equation}
\sigma_F= \sigma_{\rm rms} \Delta V \sqrt{N_{\rm ch}}
\label{eq:ch4:fluxerr}
\end{equation}
where $\sigma_{\rm rms}$ is the rms noise of the spectrum measured over a
signal-free region, $\Delta V$ is the velocity resolution in km~s$^{-1}$, and $N_{\rm ch}$ is the number of channels in the integration.
Upper limits for non-detections are given by $3\sigma_{F}$, where $N_{\rm ch}$ now
corresponds to the number of channels enclosed by the galaxy's
predicted linewidth at the 20\% peak flux level, W$_{20}$.  This
linewidth is estimated using the {\it r}-band Tully-Fisher relation from 
\citet{Kannappan13}, which is defined in terms of \HI{} profile linewidths (FWHM, or
$W_{50}$). We then estimate $W_{20}$ as $W_{50}+20\,{\rm km}\,{\rm s}^{-1}$
\citep{Haynes99,Kannappan02}.  A minimum linewidth of 40 km~s$^{-1}$ is enforced for our upper limit calculations to conservatively account for non-circular motions.  Atomic
hydrogen masses are then estimated with
\begin{equation} 
\frac{M_{\rm HI}}{\msun}=2.36 \times 10^5 \left(\frac{D}{\rm Mpc}\right)^2\left(\frac{F}{\rm Jy\,km\,s^{-1}}\right)
\end{equation}
where $D$ is the distance to the galaxy and $F$ is the measured flux \citep{Haynes84}. \textcolor{black}{For our analysis in \S\ref{sec:ch4:results}, we use indirect methods to estimate $M_{\rm HI}$ for galaxies lacking 21cm detections, but we use our upper limits to place strong constraints on the allowed values of these indirect estimates (see \S\ref{sec:ch4:indirect_gs} for further details). }

\subsubsection{(De-)Confusion}
\label{sec:ch4:deconfusion}

Over the range of distances included in the RESOLVE volume (64--100 Mpc), the physical sizes of the
GBT and Arecibo beams (FWHM) are 168--262 kpc and 66-82 kpc, so there is a
risk of source confusion in our observations. All potential cases of
confusion are automatically flagged by searching for known companions from existing redshift surveys (see \S\ref{sec:ch4:survey_def})
within twice the telescope beam FWHM and assuming all galaxies have
linewidths of \mbox{200 km s$^{-1}$} (or greater, if the linewidth has been
measured).  All automatically flagged cases are then inspected by eye
using the observed 21cm profile in conjunction with the known
redshifts and predicted linewidths of all nearby objects in order to make the
best possible judgment about whether the nearby objects are truly contributing to the \HI{} signal.  In total, approximately 14\% of our 21cm observations (or 18\% of our detections) suffer from potential
confusion with a nearby companion.  In these cases, we
constrain the 21cm flux using one of three possible approaches:
\begin{enumerate}
\item The corrected flux\textcolor{black}{, $F_c$,} is determined by summing the channels
  within the predicted $W_{50}$.  The statistical uncertainty, \textcolor{black}{$\sigma_{F_{\rm c,stat}}$}, is calculated following Eq.~\ref{eq:ch4:fluxerr}, but an additional systematic
  uncertainty, \textcolor{black}{$\sigma_{F_{\rm c,sys}}$,} is reported equaling the total flux within any
  channels overlapping multiple predicted galaxy linewidths.
\item If one half of the primary target's 21cm
  profile is judged to be uncontaminated, the flux is measured
  within the unconfused half and doubled to yield an estimate of $F_c$.  A 20\% systematic error is assigned to account for
  possible asymmetry in the 21cm profile.
\item If one half of the {\it companion} galaxy's 21cm profile is
  judged to be uncontaminated, this unconfused side is integrated,
  doubled, and subtracted from the total flux of the blended profile to obtain $F_c$.  A
  systematic uncertainty of 20\% of the companion's total flux is
  assigned to the target galaxy, again to account for possible profile
  asymmetries.  This method is not applicable if there are more than
  two potentially blended sources within the 21cm beam.
\end{enumerate}

\begin{figure}
\epsscale{1.15}
\plotone{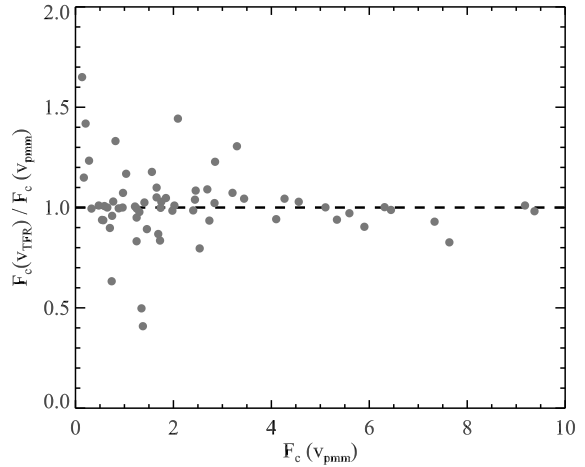}
\caption[Ratio of corrected flux using TFR-based linewidths and $V_{\rm pmm}$-based linewidths for confused sources as a function of the $V_{\rm pmm}$-based corrected flux]{Ratio of corrected flux using TFR-based linewidths, $F_c(V_{\rm TFR})$, and corrected flux using \mbox{$V_{\rm pmm}$-based} linewidths, $F_c(V_{\rm pmm})$, for confused sources as a function of the $V_{\rm pmm}$-based corrected flux. The ratios of corrected fluxes from these two methods have a median of unity and a scatter of ${\sim}$20\%.}
\label{fig:ch4:dccompare}
\end{figure}

For the deconfusion procedure, the range of heliocentric velocities subtended by each possible \HI{} source can be estimated using two possible approaches. First, $W_{50}$ may be estimated using the $r$-band Tully Fisher relation (TFR) from \citet{Kannappan13}, and then used in conjunction with estimates of the recession velocity from existing redshift surveys.  Alternatively, galaxy rotation curves from RESOLVE optical spectroscopy can be used to estimate the rotation velocity, $V_{\rm pmm}$, which is then converted
into the equivalent $W_{50}$ following equation (B6) from
\citet{Kannappan02}.  In addition to more direct measurements of rotation speed, rotation curves also typically give more reliable estimates of systemic velocities compared to single-fiber redshift surveys. However, 3-D spectroscopic observations for RESOLVE are ongoing, and at this stage $V_{\rm pmm}$ is only available
for $\sim$20\% of galaxies.  For homogeneity, we use the TFR-based linewidth predictions for all cases of confusion, but to test that the TFR-based deconfusion method is consistent with the more reliable $V_{\rm pmm}$-based method, we compare the ratio of the corrected fluxes for confused galaxies when both methods are possible. Following
\citet{Kannappan13}, we ignore any $V_{\rm pmm}$ measurements where the
rotation curve does not extend past $1.3r_{50}$ for galaxies with
morphological type earlier than Sc, where $r_{50}$ is the $r$-band half-light
radius (morphological typing for the RESOLVE survey is described in S. J. Kannappan et al., in prep.).  For types later than Sc, rotation curves extending to at
least $r_{50}$ are allowed.  These cuts avoid cases where the rotation curve does not trace the full galaxy potential.  The results of this comparison are shown in Fig.~\ref{fig:ch4:dccompare}.  We find that the two methods of deconfusion are consistent with one another, typically agreeing to within 20\% with no systematic offset. Visual inspection suggests that the largest outliers may be systems currently experiencing strong tidal interactions.  Most have rotation curve asymmetries of greater than 5\%, and some show signs of morphological disturbance.  

The goal of this procedure described above is to reliably quantify the 21cm flux and its uncertainty in cases of source confusion. Fortunately, even in the presence of confusion, a significant
fraction of 21cm observations are still useful for many analyses.
For half of the confused sources, the fluxes can be constrained to within 50\% uncertainty, and 40\% of the sources can have their fluxes constrained to within 25\%
uncertainty.  However, it is important to keep in mind that, due to the magnitude limits of existing redshift surveys, some objects may still suffer from confusion with low-mass neighbors lacking spectroscopic redshift measurements. 

\subsection{21cm Census Status and Catalog Presentation}
\label{sec:ch4:status}

\begin{figure*}
\epsscale{1.15}
\plotone{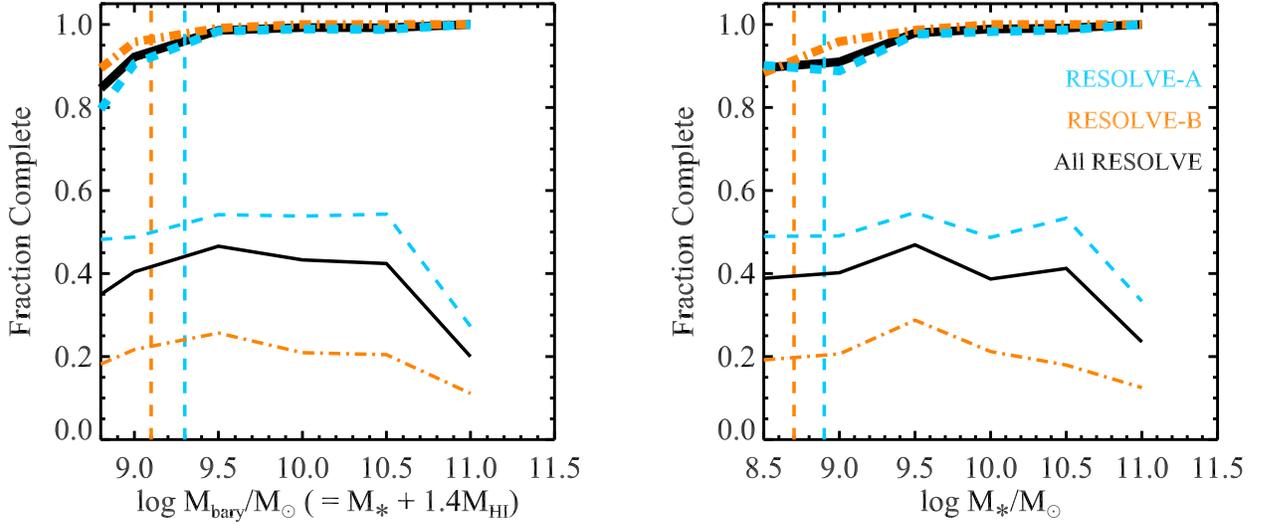}
\caption[Current 21cm census completeness as a function of baryonic and stellar mass]{Current 21cm census completeness as a function of baryonic mass (left) and stellar mass (right), shown for the full survey, as well as the separate RESOLVE-A and RESOLVE-B footprints.  The thick lines represent the current completeness levels for galaxies with integrated S/N$>$5 or $1.4M_{\rm HI}/M_{*}<0.1$.  \textcolor{black}{The thin lines represent the original completeness when just the ALFALFA survey data were available (note: ALFALFA does not cover the southern half of RESOLVE-B).}  Vertical dashed lines show the completeness limits for RESOLVE-A and RESOLVE-B.}
\label{fig:ch4:completeness}
\end{figure*}

Fig.~\ref{fig:ch4:completeness} shows the current 21cm census completeness \textcolor{black}{(where we define {\it complete} as having an \HI{} detection with S/N$>$5 or an upper limit yielding $M_{HI}/M_*<0.1$, although typically we obtain detections or limits of higher quality)}
as a function of baryonic and stellar mass (in cases where 21cm
observations are incomplete, we estimate $M_{\rm HI}$ using the
relationship between gas-to-stellar mass ratio, color, and axial ratio; \citealt{Eckert15}).  In total, the survey is
$\sim$94\% complete (94\% in RESOLVE-A, 95\% in RESOLVE-B) and is $>$85\% complete at all mass scales.

The RESOLVE 21cm catalog is available in machine-readable format in the online version of this paper.  A summary of information included in the catalog is given in Table~\ref{table:ch4:hicat}.  \textcolor{black}{The full catalog contains {\HI} data for a total of 2164 galaxies.  We include galaxies in this catalog lying within our survey volume (including a $\pm250\,{\rm km\,s^{-1}}$ buffer region; see \S\ref{sec:ch4:environment_metrics}) even if they do not fall above RESOLVE's nominal completeness limits (see \S\ref{sec:ch4:sample}).} Additional extracted quantities (linewidths, systemic velocities, asymmetries) will be included in future publications. 

\begin{table*}
\caption{RESOLVE 21cm Catalog Description}
\label{table:ch4:hicat}
\begin{center}
\begin{tabular}{@{}cl}
\hline\hline
Column & Description \\
\hline
1 & RESOLVE Designation \\
2 & R. A. \\
3 & Decl. \\
4 & Source of \HI{} data \\
5 & Total 21cm flux, $F$ \\
6 & Uncertainty on total 21cm flux, $\sigma_F$ \\
7 & rms noise of the observed spectrum assuming 10 km s$^{-1}$ channels, $\sigma_{\rm rms}$\\
8 & Flag indicating total 21cm flux is an upper limit \\
9 & Flag indicating if the \HI{} source is confused \\
10 & 21cm flux corrected for source confusion, $F_c$ \\
11 & Statistical uncertainty on confusion-corrected 21cm flux, $\sigma_{\rm F,stat}$  \\
12 & Additional systematic uncertainty on confusion-corrected 21cm flux, $\sigma_{\rm F,sys}$ \\
13 & Method used to determine the confusion-corrected flux and its systematic error \\

\hline
\end{tabular}
\end{center}
\end{table*}

\subsection{Environment Metrics}
\label{sec:ch4:environment_metrics}
\textcolor{black}{To characterize the environments of galaxies, we use group identifications with corresponding dark matter halo masses and central/satellite designations (\S\ref{sec:ch4:mhalo}), large-scale structure densities (\S\ref{sec:ch4:dmdens}), and the distance to the nearest massive group (\S\ref{sec:ch4:group_sep}).} Environment metrics can become unreliable in close proximity to survey edges, so to help minimize this issue, RESOLVE has a buffer region extending $\pm 250$ km s$^{-1}$ from the nominal survey $V_{\rm LG}$ range of 4500 to 7000 km s$^{-1}$.   \textcolor{black}{Additionally, the RESOLVE-A volume is embedded within the much larger ECO (Environmental COntext) catalog \citep{Moffett15}. ECO provides a ${>}10\times$ larger volume over a slightly larger redshift range, \mbox{$V_{\rm LG}=2530-7470\,{\rm km\,s^{-1}}$}, with a completeness limit roughly equivalent to that of RESOLVE-A. ECO is compiled from the same list of redshift catalogs as RESOLVE (see \S\ref{sec:ch4:survey_def}). All environment metrics for RESOLVE-A are calculated using ECO.  RESOLVE-B is not embedded within a larger redshift survey of comparable completeness and is more subject to edge effects.  However, we have accounted for potential biases due to edge effects (see \S\ref{sec:ch4:dmdens} and \S\ref{sec:ch4:group_sep}) and find that our results are not sensitive to the inclusion of the affected galaxies (\S\ref{sec:ch4:dmdens_gs} and \S\ref{sec:ch4:gs_groupdist}).}

All environment metrics described in the following sections are available in a machine-readable table included in the online version of this paper.  A description of each column is provided in Table~\ref{table:ch4:envcat}.  

\begin{table*}
\begin{center}
\begin{threeparttable}
\caption{Environment Metrics}
\label{table:ch4:envcat}
\begin{tabular}{@{}cl}
\hline\hline
Column & Description \\
\hline
1 & RESOLVE Designation \\
2 & Group ID ($M_*$-limited sample)\\
3 & Group dark matter halo mass, $M_h$ ($M_*$-based HAM)\\
4 & Group ID ($M_{\rm bary}$-limited sample)\\
5 & Group dark matter halo mass $M_h$ ($M_{\rm bary}$-based HAM)\\
6 & Large-scale structure density, $\rho_{\rm LSS}$ ($M_*$-based HAM)\\
7 & Large-scale structure density corrected for edge-effects where necessary ($M_*$-based HAM)\\
8 & Large-scale structure density, $\rho_{\rm LSS}$ ($M_{\rm bary}$-based HAM)\\
9 & Large-scale structure density corrected for edge-effects where necessary ($M_{\rm bary}$-based HAM)\\
10 & Distance to nearest group of $M_h>10^{12}\,\msun$, ${D_{\rm NG}}^a$ ($M_*$-based HAM) \\
11 & Flag indicating $D_{\rm NG}$ may be unreliable due to proximity to survey edge ($M_*$-based HAM)\\
12 & Distance to nearest group of $M_h>10^{12}\,\msun$, ${D_{\rm NG}}^a$ ($M_{\rm bary}$-based HAM) \\
13 & Flag indicating $D_{\rm NG}$ may be unreliable due to proximity to survey edge ($M_{\rm bary}$-based HAM)\\
\hline
\end{tabular}
\begin{tablenotes}
\item $^a$ Only calculated for $M_h<10^{12}\,\msun$ groups.
\end{tablenotes}
\end{threeparttable}
\end{center}
\end{table*}

\subsubsection{Group Dark Matter Halo Masses}
\label{sec:ch4:mhalo}

Dark matter halo masses serve as a fundamental way to characterize galaxy groups, and they likely play a key role in galaxy evolution (see \S\ref{sec:ch4:intro}).  To assign group halo masses, we first identify galaxy groups using the friends-of-friends (FoF) technique described in \citet{Berlind06}. Group dark matter halo masses ($M_h$) are then estimated using halo abundance matching (HAM; \citealt{Peacock00}, \citealt{Berlind02}), where we assume a monotonic relationship between the integrated stellar mass of a group and its dark matter halo mass, then assign masses by matching the cumulative abundance of groups at each integrated stellar mass to the cumulative theoretical group dark matter halo mass function of \citet{Warren06}.  Note that  we are not assigning masses to dark matter subhalos, so by definition all galaxies in a group share the same dark matter halo mass.  

The relative simplicity of the FoF/HAM method makes it advantageous for estimating halo masses, but it carries with it several potential sources of error.  First, the FoF algorithm can blend or fragment true groups, which then affects the overall completeness and reliability of identified groups. There is no single choice of FoF linking lengths that completely avoids both of these problems simultaneously. Cosmic variance is another potential source of error.  Optimized linking lengths are typically determined from large mock catalogs and expressed in terms of the mean particle density of the volume.  Therefore, group identifications may be influenced by cosmic variance if the volume in question is not large enough such that its average galaxy number density is significantly higher or lower than average.  HAM can likewise suffer from cosmic variance in the sense that the abundances of groups at different masses may be biased if the volume in question is not large enough.  Additionally, the parameter  used to predict halo mass (typically total group stellar mass, but alternatives include total group luminosity or total group baryonic mass) can potentially build in apparent correlations between galaxy properties and halo mass that are actually correlations between galaxy properties and the parameter used to predict halo mass (see \S\ref{sec:ch4:mhalo_gs} and \S\ref{sec:ch4:lss_mhalodef} for detailed discussions of this issue).  As implemented here, HAM also ignores any intrinsic scatter around the relationship between halo mass and the parameter used to estimate it. 

\textcolor{black}{For this work, the line-of-sight and plane-of-sky linking lengths, $b_{\|}$ and $b_{\bot}$, are set to 0.07 and 1.1 times the mean inter-galaxy spacing, $l={n}^{-1/3}$, where $n$ is the mean galaxy number density in the volume.}  These values are chosen based on the recommendations of \citet{Duarte14} for environmental studies of galaxies, and are separately confirmed in \citet{Eckert16} as ideal linking lengths to minimize blending of low-$N$ groups and improve recovery of galaxies with high peculiar velocities.  For this choice of linking lengths, \citet{Duarte14} quantify the level of fragmentation (fraction of true groups broken into two or more groups by the FoF algorithm), merging (two or more true groups blended into a single group by the FoF algorithm), completeness (fraction of galaxies in a true group recovered in the FoF-identified group), and reliability (fraction of objects in an FoF-identified group that are truly part of that group).  In true group dark matter halos with masses of $10^{12-13}\,\msun$, between 10\% and 20\% of groups suffer from fragmentation, and a similar fraction suffer from merging.  However, the estimated groups have high completeness ($>$95\%) and reliability (90--95\%).  With these linking lengths, the quality of the estimated groups tends to decline as halo mass increases.  For halos with masses of $10^{13-14}\,\msun$, the merging and fragmentation rates increase by $\sim$10\%, while the completeness and reliability decrease by $\sim$5--10\%.  \citet{Duarte14} do not quantify the quality of FoF group identification at the lower halo masses (${\sim}10^{11-12}\,\msun$) that dominate our sample, although given that the group quality tends to increase with decreasing halo mass, we expect the quality of groups in the ${\sim}10^{11-12}\,\msun$ regime to be at least comparable to the $10^{12-13}\,\msun$ regime. \citet{Moffett15} use mock catalogs to quantify the typical error on the halo masses estimated from HAM with our choice of linking lengths and find typical random uncertainties of 0.12 dex, although errors can be significantly larger where groups suffer from merging or fragmentation\footnote{\citet{Moffett15} also find that halo masses below $10^{12}\,\msun$ are systemically overestimated by $\sim0.15$ dex on average by the FoF/HAM procedure.  However, \citet{Eckert16} determine that this apparent offset is due to different overall densities of the mock catalog used to quantify uncertainties and the ECO catalog itself. Using a mock catalog specifically chosen to match the density of ECO shows no offset between true and estimated halo masses obtained from abundance matching.  Therefore, we apply no offset to the halo mass scale in this paper. }.

\textcolor{black}{Due to the different completeness limits and volume sizes, groups in RESOLVE-A and RESOLVE-B are identified in slightly different ways.  For RESOLVE-A, groups are found by running the FoF algorithm on the larger ECO sample with a stellar mass completeness limit of $M_*>10^{8.9}\,\msun$ (the mean inter-galaxy spacing for this sample is $2.8\,\hmpc$). Identifying groups using ECO, which has a \mbox{${>}10$ times} larger volume than \mbox{RESOLVE-A}, helps to minimize bias caused by cosmic variance.  For RESOLVE-B, which is $\sim$40 times smaller than ECO, identifying groups using the linking lengths determined from the mean inter-galaxy spacing in this volume ($2.5\,\hmpc$) could be highly subject to the effects of cosmic variance (especially because RESOLVE-B is thought to be overdense; see \citealt{Moffett15}, \citealt{Eckert16}, and \S\ref{sec:ch4:cosmic_variance}).  Instead, we apply the same physical linking lengths determined using ECO for RESOLVE-A to a version of RESOLVE-B with the stellar mass completeness limit matched to RESOLVE-A. Abundance matching is used to estimate group halo masses in ECO for RESOLVE-A, and again to avoid bias due to cosmic variance, we fit a spline to the resulting $M_*-M_h$ relation in ECO and use this fit to assign halo masses to RESOLVE-B.}

Halo masses based on integrated group stellar mass are used by default in this paper, but we will also use group halo masses estimated from integrated baryonic mass.  Quantitatively, the process of estimating halo masses via baryonic mass is identical to the description above, except we use the alternate completeness limits given in \S\ref{sec:ch4:survey_def}.  Halo masses for groups with centrals that lie below the nominal mass completeness limits are determined by downward extrapolation of the stellar (or baryonic) mass-halo mass relationship determined from the mass-limited ECO sample, although galaxies below the mass completeness limits are not incorporated into the analysis in this paper.

Although ECO suffers from cosmic variance less than the RESOLVE volumes, it is not itself necessarily free from bias.  We attempt to quantify the potential size of the offset in ECO's halo mass function due to cosmic variance using the results of \citet{Hu03} who quantify the potential error in number counts based on a volume size and mass limit.  Extrapolating Fig.~2 from \citet{Hu03} down to a mass limit of $\sim10^{11}\,\msun$ (comparable to the minimum halo mass in ECO) and using a radius of 36 $\hmpc$ (determined by treating ECO as a sphere with volume \mbox{192369.3 h$^{-3}$ Mpc$^{3}$),} we estimate that ECO's halo mass function may be biased by ${\sim}\pm0.1$ dex, which translates into a comparable uncertainty in our halo mass scale.   An alternative calibration of cosmic variance by \citet{Trenti08} also yields an estimated potential bias of ${\sim}\pm$0.1 dex in the halo mass function.  A more robust estimation of uncertainties from cosmic variance specifically for RESOLVE/ECO is in preparation (J. Cisewski et al, in preparation).

\begin{figure}
\epsscale{1.15}
\plotone{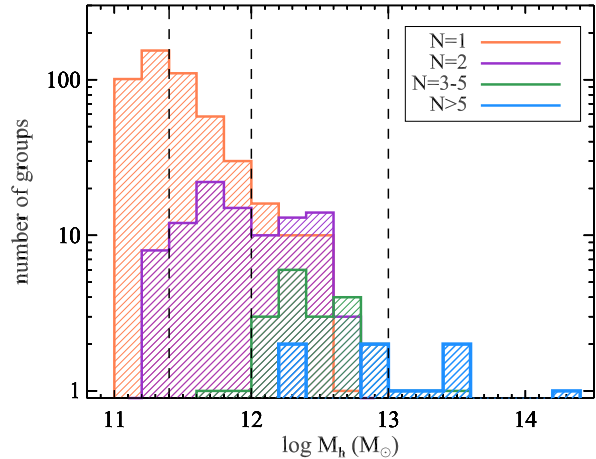}
\caption[Distribution of group halo masses for groups with different numbers of members]{Distribution of group halo masses for groups with different numbers of members ($N$).  Vertical dashed lines show characteristic group halo mass scales used often in this work, $M_h=10^{11.4}\,\msun$, $10^{12}\,\msun$, and $10^{13}\,\msun$. Singleton (N=1) groups are common at low halo mass.}
\label{fig:ch4:groupn_mhalo_hist}
\end{figure}

Throughout this work, we consider the most massive galaxy to be the ``central" galaxy of a group.  We also refer to galaxies with no satellites as ``centrals."  These singleton groups preferentially exist at group halo masses $<10^{11.4}\,\msun$, as seen in Fig.~\ref{fig:ch4:groupn_mhalo_hist} which shows the distribution of group halo masses for groups with different numbers of members.  

\subsubsection{Cosmic Web Density} 
\label{sec:ch4:dmdens}

The mass density of the cosmic web beyond the group scale serves as a way to parameterize the larger-scale environments of galaxies.  \citet{Carollo13} give a thorough assessment of the advantages and disadvantages of estimating the density field using $N$th nearest-neighbors, fixed apertures, or Voronoi tessellations.  Following their arguments, we characterize the large-scale density around each group using total projected mass density within the distance to the third-nearest {\it group} (not galaxy).  Specifically, we define this as
 \begin{equation}
 \rho_{\rm LSS} = \frac{1}{\pi R_3^2}\sum\limits_{i=0}^3 M_{h,i}
 \end{equation}
where M$_{h,i}$ are the group halo masses and $R_3$ is the projected distance to the third-nearest group.  We only consider projected distances to groups with recession velocity differences of $<$500 km~s$^{-1}$.  This relative velocity criterion is commonly used in the literature to select neighboring galaxies, but we also employ mock catalogs to confirm that the vast majority of neighboring groups also have relative velocities less than this value.  

Using the $N$th nearest group has two key advantages over using the $N$th nearest galaxy.  First, it minimizes the correlation between the density metric and group halo mass (although the correlation is not completely removed).  Second, $N$th nearest galaxy density estimates change from reflecting a group density for cases where the number of group members is greater than $N$, to reflecting an intergroup density for cases where the number of group members is less than $N$.  \citet{Carollo13} show that using the $N$th nearest group instead of the $N$th nearest galaxy provides a more consistent large-scale structure density estimator.  Also like \citet{Carollo13}, we find little difference between different choices of N, finding that $N=3$ and $N=5$ yield consistent densities.  We opt to use $N=3$ because it minimizes the fraction of groups whose density estimate is compromised by proximity to the survey edge.

Densities  may be underestimated when the distance to the third-nearest group is larger than the distance to the edge of the survey volume.  In these cases, we follow the method of \citet{Kovac10} and correct the densities by dividing them by the fraction of the projected area within $R_3$ that lies within the survey volume.  Typical corrections are modest, changing densities by less than a factor of 2.  For groups near the edges of RESOLVE-A, we use the larger ECO volume to calculate densities, so only  6\% of groups in RESOLVE-A require corrections.  However, since RESOLVE-B is not embedded within a larger survey of equal depth, and is a very thin volume, 60\% of its groups have density estimates that require corrections.  \textcolor{black}{Despite this large fraction, the generally small magnitude of the density corrections means edge effects do not strongly compromise our results (see \S\ref{sec:ch4:dmdens_gs} for further discussion).}

Our chosen density estimator ignores any mass not contained within halos and is not meant to be used as an estimate of the {\it true} cosmic web density in a given region.  However, this metric provides a means to compare the {\it relative} large-scale densities throughout our survey volume.  Therefore, we express all densities as a multiple of the median density measured within our volume, rather than units of $\msun\,{\rm Mpc}^{-2}$. 

\begin{figure}
\epsscale{1.15}
\plotone{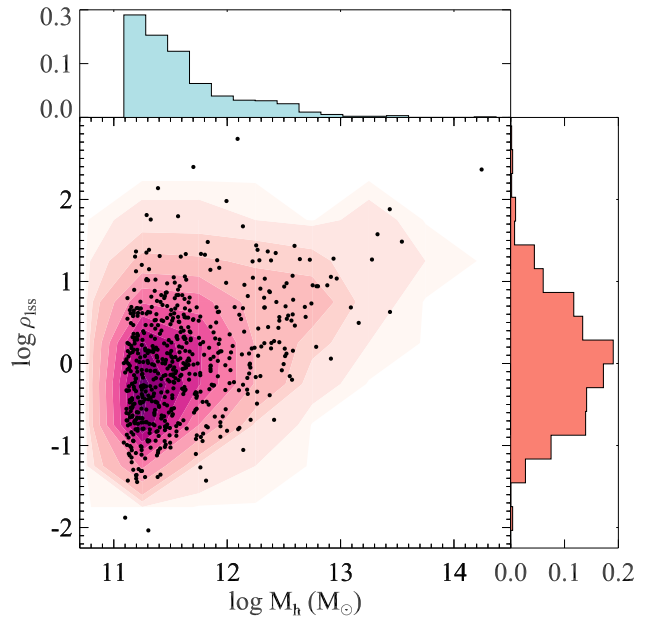}
\caption{\textcolor{black}{Large-scale structure density ($\rho_{\rm LSS}$) versus group dark matter halo mass ($M_h$). $\rho_{\rm LSS}$ is expressed as a multiple of the median density measured within our volume. Points represent individual groups with contours highlighting the distribution of points for the \mbox{$M_*>10^{8.9}\,\msun$} limited sample.  Histograms on the top and right panels show the relative fractions of groups as a function of $\rho_{\rm LSS}$ and $M_h$.  We include groups whose $\rho_{\rm LSS}$ estimates require corrections for edge effects, and those corrections have been applied.}}
\label{fig:ch4:mhalo_dmdens_contour}
\end{figure}

\textcolor{black}{Fig.~\ref{fig:ch4:mhalo_dmdens_contour} shows the distribution of $M_h$ and $\rho_{\rm LSS}$ for our final stellar mass-limited sample (see \S\ref{sec:ch4:sample}).  Importantly, at fixed halo mass, particularly below $M_h=10^{12}\,\msun$, groups span a wide range of $\rho_{\rm LSS}$ (also seen by \citealt{Carollo13}) allowing an analysis of how large-scale environment affects gas content independent of halo mass.}

\subsubsection{Distance to Nearest Massive Group}
\label{sec:ch4:group_sep}

\textcolor{black}{Studies highlighting the possible effect of group-group interactions (e.g., flyby interactions, competitive gas accretion; \citealt{Wetzel12,Hearin16})  suggest the physical separations between groups can have an important impact on their evolution.  Therefore, as a third environmental parameter, we estimate the distance of each group to its closest neighboring group, $D_{\rm NG}$, defined as the group with the smallest projected separation and recession velocity difference $<$500~km~s$^{-1}$.  $D_{\rm NG}$ estimates are normalized by the virial radius, $R_{\rm vir}$, of the nearest group's dark matter halo (where we define $R_{\rm vir}$ as $R_{200m}$, i.e., the radius where the matter density of the halo is 200 times the universal mean matter density). Although $D_{\rm NG}$ can be estimated independently of group mass, our analysis specifically focuses on the distance of $M_h<10^{11.4}\,\msun$ groups to $M_h>10^{12}\,\msun$ groups.  The motivation for this choice and details of the analysis are discussed in \S\ref{sec:ch4:gs_groupdist}.}

\textcolor{black}{Estimates of $D_{\rm NG}$ can be affected by a number of uncertainties. First, the FoF algorithm used to identify groups can misclassify centrals and satellite galaxies.  To account for this issue, we estimate the rate of blending/fragmentation as a function of group separation using mock catalogs, limiting our analysis to mock catalogs with mean number densities within 20\% of the ECO volume (0.023 ${\rm Mpc^{-3}}$).  For our choice of linking lengths, blending is relatively negligible compared to fragmentation, which is primarily an issue at small group separations.  $D_{\rm NG}$ can also be unreliable when the measured value is less than the distance to the edge of the survey volume (including the buffer regions).  In these cases we can still place limits on the possible values of $D_{\rm NG}$.  The lower limit is estimated by assuming a $M_h=10^{15}\,\msun$ halo resides just outside the edge of the volume.  The upper limit of $D_{\rm NG}$ is the currently measured value.   The impact of these uncertainties on $D_{\rm NG}$ is discussed further in \S\ref{sec:ch4:gs_groupdist}.}

\subsection{Definition of Mass-limited Samples}
\label{sec:ch4:sample}

\textcolor{black}{Unless stated otherwise, all analyses presented in \S\ref{sec:ch4:results} use a stellar mass limited sample with \mbox{$M_{*} > 10^{8.9}\,\msun$} corresponding to the estimated stellar mass completeness limit of RESOLVE-A (but see \S\ref{sec:ch4:completeness_corrections}). Although RESOLVE-B has a completeness limit of $M_*=10^{8.7}\,\msun$, we do not include these additional lower-mass galaxies in our main analysis in order to have a sample with uniform depth. However, in \S\ref{sec:ch4:cosmic_variance} we discuss an analysis of just RESOLVE-B down to its true completeness limit. For the full sample, our $M_*>10^{8.9}\,\msun$ selection yields a total of 941 galaxies, 636 in RESOLVE-A and 305 in RESOLVE-B (there are 373 galaxies in RESOLVE-B when limited to $M_*>10^{8.7}\,\msun$).} 

Although RESOLVE was originally designed to be complete in baryonic mass, a stellar mass-limited selection is our default for this study.  Many environmental processes that remove gas, such as ram-pressure or viscous stripping, most directly affect the gas content of a galaxy, not the stellar content. Therefore, when examining which environments host gas removal processes, it is most intuitive to compare gas content at fixed {\it stellar} mass. The situation is more complicated for starvation, which implies reduced star formation and thus coupled gas and stellar mass deficiency, and tidal interactions between galaxies, which can alter both gas and stellar content of a system simultaneously.   The default stellar mass-selected approach employed in this study tends to highlight gas removal interpretations at the expense of starvation interpretations.   \textcolor{black}{In \S\ref{sec:ch4:mhalo_gs} and \S\ref{sec:ch4:lss_mhalodef}, we discuss how our results vary if we use a baryonic mass-limited sample, defined as $M_{\rm bary}>10^{9.3}\,\msun$.}

\subsubsection{Indirect Gas Mass Estimates}
\label{sec:ch4:indirect_gs}

The \HI{} census contains a number of upper limits or confused detections, leading to uncertainty in total gas content.  However, our efforts to obtain strong limits and deconfuse blended profiles allow us to place strong constraints on gas masses in most of these situations. \textcolor{black}{To estimate true gas-to-stellar mass ratios (defined as $G/S=M_{\rm HI}/M_*$) in these cases, we combine the probability distribution of $G/S$ as a function of color and axial ratio (see Figs.~13 and 14 from \citealt{Eckert15}) with additional information based on measured limits or deconfusion procedures}. Specifically:
\begin{itemize}
\item For upper limits, a value is drawn randomly from the $G/S$ probability distribution, but we set the probability to zero above the measured upper limit value.  
\item For confused detections with $\sigma_{F_{c,sys}}/F_c < 0.25$ (i.e., confused but with relatively small additional uncertainty) we adopt the confusion-corrected $G/S$.
\item For confused detections with $\sigma_{F_{c,sys}}/F_c > 0.25$, a value is drawn randomly from the $G/S$ probability distribution, but the probability is set to zero below $M_{{\rm HI},c} - \sigma_{M_{\rm HI},sys}$ and above $1.05 \times M_{{\rm HI},c}$.  This lower bound represents the absolute minimum possible flux of the confused detection (only the flux from unconfused channels in the spectrum), while the upper bound accounts for the typical amount of flux missed in the wings of a profile when integrating from $\pm W_{50}/2$.
\end{itemize}
As previously mentioned, we ignore the contribution from H$_2$ in our total gas budget, but expect it to be negligible for the majority of our sample.

\subsubsection{Completeness Corrections} \label{sec:ch4:completeness_corrections}
\textcolor{black}{We consider RESOLVE-B to be a 100\% complete data set (see \S3.6 of \citealt{Eckert16}), and we can therefore use it to construct empirical completeness corrections for RESOLVE-A. We follow the methodology described in \citet{Moffett15} and \citet{Eckert16}, who compared two-dimensional galaxy number density fields in the space of $M_r$ vs.\ $g-i$ color for the SDSS DR7, ECO, and RESOLVE-B samples to derive survey completeness correction fields referenced to \mbox{RESOLVE-B}. Here, the relevant completeness correction field is simply the RESOLVE-B field divided by the RESOLVE-A field. Instead of determining the completeness correction field as a function of $M_r$ and $g-i$ color, in this work, we use $M_*$ and $g-i$ for our stellar mass-limited sample and $M_{\rm bary}$ and $g-i$ for our baryonic-mass limited sample. This analysis results in multiplicative correction factors that are used to weight RESOLVE-A galaxies when analyzing galaxy property distributions. The median correction factor in RESOLVE-A is $\sim$1.1, with no corrections larger than 1.2. By definition, the correction factors in RESOLVE-B are all 1.0. Although we incorporate completeness corrections throughout our analysis, they do not have any impact on our results.}

\section{Results}
\label{sec:ch4:results}

In the following section, we present our findings on the relationship between galaxy gas fraction and environment on multiple scales. First, we investigate the influence of group halo mass, specifically whether galaxies in intermediate-mass group halos show signatures of gas deficiency similar to those seen in massive groups and clusters (\S\ref{sec:ch4:mhalo_gs}).  Next, we explore whether the large-scale density of the cosmic web and the proximity of the nearest significantly larger group affect gas content independent of halo mass (\S\ref{sec:ch4:lss_gs}). \textcolor{black}{We conclude by examining whether our findings are affected by cosmic variance by comparing the results from RESOLVE-A and RESOLVE-B separately (\S\ref{sec:ch4:cosmic_variance}).} Throughout our analysis, we  often separate  central and satellite galaxies since environment may affect these subpopulations in different ways. 

\subsection{Group Halo Mass}
\label{sec:ch4:mhalo_gs}

\begin{figure*}
\epsscale{1.15}
\plotone{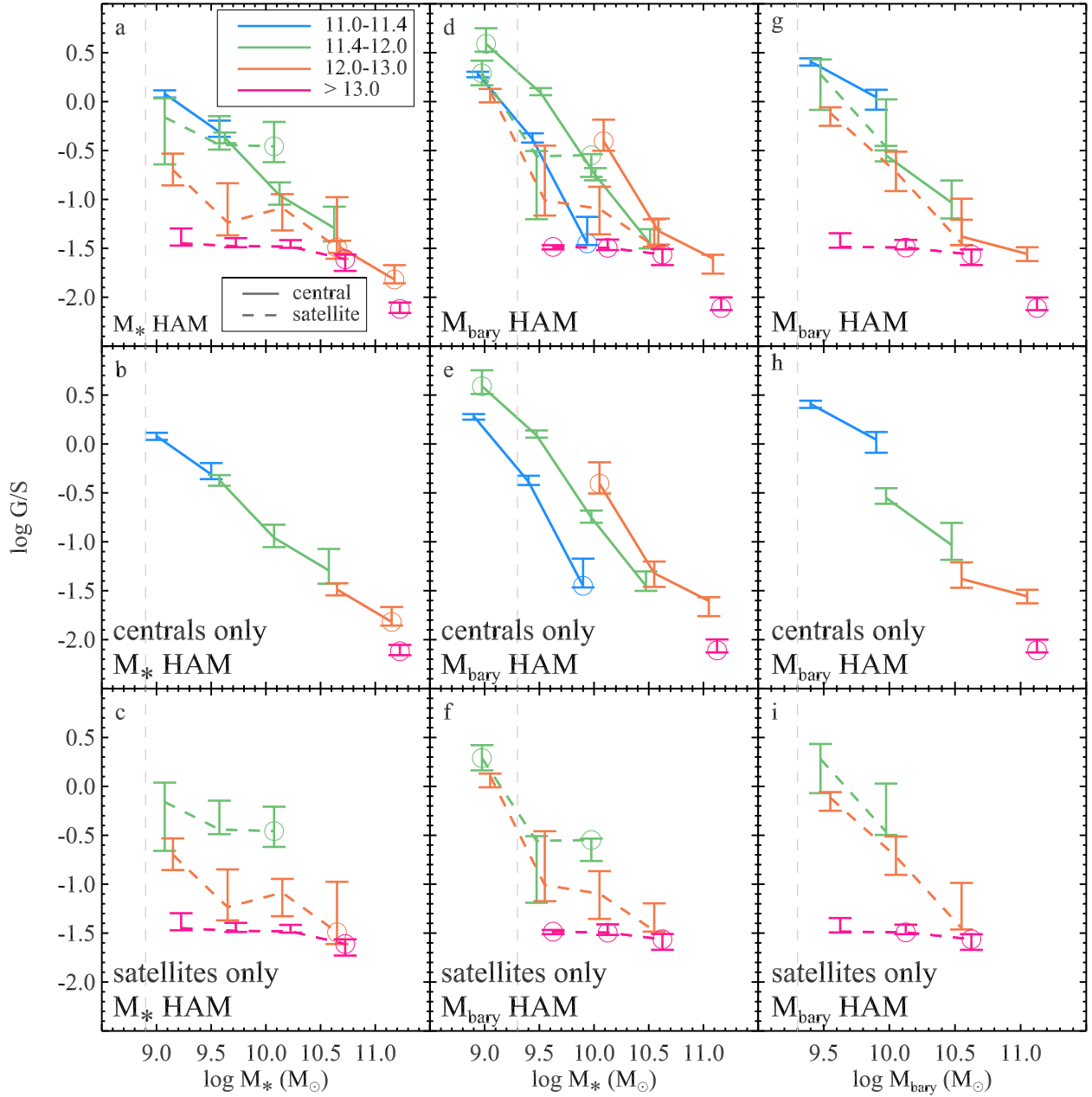}
\caption{{\bf(a)} Median $G/S$ versus $M_*$ as a function of group halo mass estimated from halo abundance matching using integrated group stellar mass. Centrals and satellites are plotted separately using solid and dashed lines. {\bf(b)} Same as panel (a), but only central galaxies are shown. {\bf (c)} Same as panel (a), but only satellite galaxies are shown. {\bf (d,e,f)} Same as panels a--c, but now halo masses are estimated via halo abundance matching using integrated group baryonic mass.  {\bf (g,h,i)} Same as panels d--f, but the x-axis variable is now baryonic mass.  All plots require at least 5 galaxies per bin, and points with open circles indicate bins with fewer than 20 galaxies, which may have unreliable error bars from bootstrap resampling.  The vertical dashed lines represent the mass completeness limits of the plotted samples.  Individual points in the same bin are offset slightly for clarity. The plotted relationships for central galaxies are highly dependent on the group parameter used in halo abundance matching, as are the relative relationships between centrals and satellites.  However, satellites show consistent behavior in all three plots: $G/S$ systematically decreases as halo mass increases, indicating group processes which lower gas content in satellites are present at moderate halo masses as low as $10^{12}\,\msun$.}

\label{fig:medgs_mstars_mbary_3panel}
\end{figure*}

To understand how group halo mass drives variations in the relationship between gas content and stellar mass, Fig.~\ref{fig:medgs_mstars_mbary_3panel}a shows median G/S as a function of stellar mass in different group halo mass regimes, separated into central and satellite galaxies. Fig.~\ref{fig:medgs_mstars_mbary_3panel}b and Fig.~\ref{fig:medgs_mstars_mbary_3panel}c show the same data, but with the centrals and satellites plotted separately for clarity.  Uncertainties on the medians in each bin are determined from bootstrap resampling (10,000 resamples with replacement) of the data
and reflect the 68\% confidence interval on the median. The bootstrap
assumes the observed distribution of data is a decent estimate of the
true distribution, but this assumption can break down when few data
points are available \citep{Chernick08}. We employ the ``smoothed" bootstrap, where for each data point in the bootstrap resample, $x_i$, we add random noise drawn from the normal distribution $N(x_i,\sigma^2)$, where $\sigma=s/\sqrt{N}$, $s$ is the usual sample standard deviation,
and $N$ is the sample size \citep{Hesterberg04}. Adding this small amount of noise reduces the
discreteness of the resulting bootstrap distribution of the median that can arise with small
samples sizes. Nonetheless, we only plot bins with at least five points,
and we are cautious about interpreting any bins with less than 20
points, which we have marked with open circles (in Fig.~\ref{fig:medgs_mstars_mbary_3panel} and all subsequent figures). The stellar mass completeness limit is
shown in Figs.~\ref{fig:medgs_mstars_mbary_3panel}a--c by the gray dashed line.

At fixed stellar mass, satellites have systematically lower $G/S$ as halo mass increases.  Meanwhile, centrals follow a smooth
relationship between $G/S$ and stellar mass with no secondary dependence
on group halo mass, implying that halo mass and central stellar mass
are closely linked.  However, we stress that the close link between
central stellar mass and group halo mass is a built-in result; $M_h$ is
itself estimated by assuming a monotonic relationship with integrated group stellar mass that has zero scatter, and the group stellar mass is typically dominated by the central galaxy (at least in groups below $M_h\sim10^{13}\,\msun$, which make up the vast majority of our sample).

To further illustrate how correlations between galaxy properties and
 halo mass can be manufactured, we re-examine the $G/S$-$M_*$-$M_h$ relationship using halo
 masses derived from HAM based on total group {\it baryonic} mass rather than total group stellar
 mass.  For this analysis, we use the baryonic mass-limited subset of
 RESOLVE with $M_{\rm bary}>10^{9.3}\,\msun$ (which also represents the
 effective stellar mass completeness limit for this subsample) \textcolor{black}{containing 767 galaxies in RESOLVE-A and 310 galaxies in RESOLVE-B for a total of 1077 galaxies.}  Halo masses estimated using integrated baryonic mass (uncommon in the literature) yields similar results to those determined using $r-$band luminosity (common in the
 literature) due to the close correlation between $r$-band luminosity and baryonic mass, notably closer than between $r$-band luminosity and stellar mass \citep{Kannappan13}. 

The new $G/S$ vs.\ $M_*$ relationships with $M_h$ estimated using baryonic mass are shown in Figs.~\ref{fig:medgs_mstars_mbary_3panel}d--f. There is no longer a smooth
 relationship between $G/S$ and $M_*$ for centrals, but rather a secondary
 dependence on halo mass such that at fixed stellar mass, centrals
 with higher G/S fall into higher-mass halos.  Again, this behavior
can be understood as a consequence of defining group halo mass in terms of integrated baryonic mass.  At fixed
 stellar mass, galaxies with higher G/S will have higher baryonic
 masses.  Therefore, by definition, they will be assigned higher halo
 masses. 
 
It is possible to recover a smooth correlation for centrals with this alternative halo mass definition.  In \mbox{Figs.~\ref{fig:medgs_mstars_mbary_3panel}g--i} we show
 $G/S$ vs $M_{\rm bary}$ (instead of $M_*$) broken up by group halo mass, where
 group halo masses are again based on the integrated baryonic mass.
 These plots are analogous to Figs.~\ref{fig:medgs_mstars_mbary_3panel}a--c in that the group halo masses are
 based on the variable on the x-axis, and the behavior of Figs.~\ref{fig:medgs_mstars_mbary_3panel}g--i is
 qualitatively similar to Fig.~\ref{fig:medgs_mstars_mbary_3panel}a--c.  In particular, the $G/S$ vs.\ $M_{\rm bary}$
 relation for centrals in Fig.~\ref{fig:medgs_mstars_mbary_3panel}h is more smooth, like the $G/S$ vs.\ $M_*$ relation for centrals in Fig.~\ref{fig:medgs_mstars_mbary_3panel}b, although there are discontinuities between different halo mass regimes at fixed $M_{\rm bary}$. A possible explanation for these discontinuities is that
 the centrals tend to account for a larger fraction of the integrated stellar mass
 than they do the integrated baryonic mass, leading to a stronger relation
 between $M_h$ and central $M_*$ compared to $M_h$ and central $M_{\rm bary}$.

As we have argued, Fig.~\ref{fig:medgs_mstars_mbary_3panel} illustrates the caution that must be taken when
interpreting relationships between galaxy properties and group halo
masses determined via HAM.  The built-in biases of HAM limit the conclusions we can draw.  Nonetheless, we are able to identify some consistent behavior among satellite galaxies regardless of how halo mass is estimated.  For satellites at fixed stellar or baryonic mass, $G/S$ progressively decreases as $M_h$ increases, implying that group processes that lower satellite gas content have a larger impact in more massive group halos.  Using the satellites in the lowest halo mass regime where they are available ($M_h=10^{11.4-12}\,\msun$; satellites in lower-mass halos are extremely rare in our sample) as the reference to compare to satellites at higher halo mass, there is evidence for systematically lower $G/S$ in satellites within groups down to $M_h=10^{12}\,\msun$, although in Fig.~\ref{fig:medgs_mstars_mbary_3panel}c, only the $M_h>10^{13}\,\msun$ satellites show statistically significant lower $G/S$ below $M_* \sim 10^{9.5}$. In Fig.~\ref{fig:medgs_mstars_mbary_3panel}f, the gas deficiency down to $M_h=10^{12}\,\msun$ is at a marginal level around $M_*\sim10^{9.5}\,\msun$ at least partly due to the small number of satellites in $M_h=10^{11.4-12}\,\msun$ halos under the baryonic mass-limited selection.  

The behavior of satellites {\it relative to centrals} is not as consistent.  In Fig.~\ref{fig:medgs_mstars_mbary_3panel}a, satellites with $M_*<10^{10}\,\msun$ in halos down to at least $M_h=10^{12}\,\msun$ have $G/S$ below all centrals with the same stellar mass, with a hint of a similar result down to $M_h=10^{11.4}\,\msun$.  However, in Fig.~\ref{fig:medgs_mstars_mbary_3panel}d, satellites no longer fall systematically below all centrals. Comparing $G/S$ of satellites to centrals is complicated by the fact that the behavior of central galaxies is strongly affected by the built-in biases from the choice of the integrated quantity used in HAM.  Furthermore, as we will discuss in \S\ref{sec:ch4:dmdens_gs} and \S\ref{sec:ch4:gs_groupdist}, centrals may themselves become gas deficient due to processes associated with the larger-scale environment (in turn altering HAM halo mass estimates that are based on integrated group baryonic mass).  Therefore, assessing halo mass scales associated with gas deficiency by comparing gas fractions of satellites with those of centrals may not always be appropriate when using HAM.

Despite the complexities of comparing gas fraction, stellar mass, and group halo mass for centrals and satellites, we can draw conclusions about the influence of group environment on satellite gas fractions: there is very strong evidence for gas deficiency in $M_h>10^{12}\,\msun$ satellites, although this deficiency is not definitive in the lowest stellar mass regime of \mbox{$M_h=10^{12-13}\,\msun$} groups.

\subsubsection{The $G/S$ Versus $M_h$ relation}
\label{sec:gs_vs_mhalo}

\begin{figure*}
\epsscale{1.15}
\plottwo{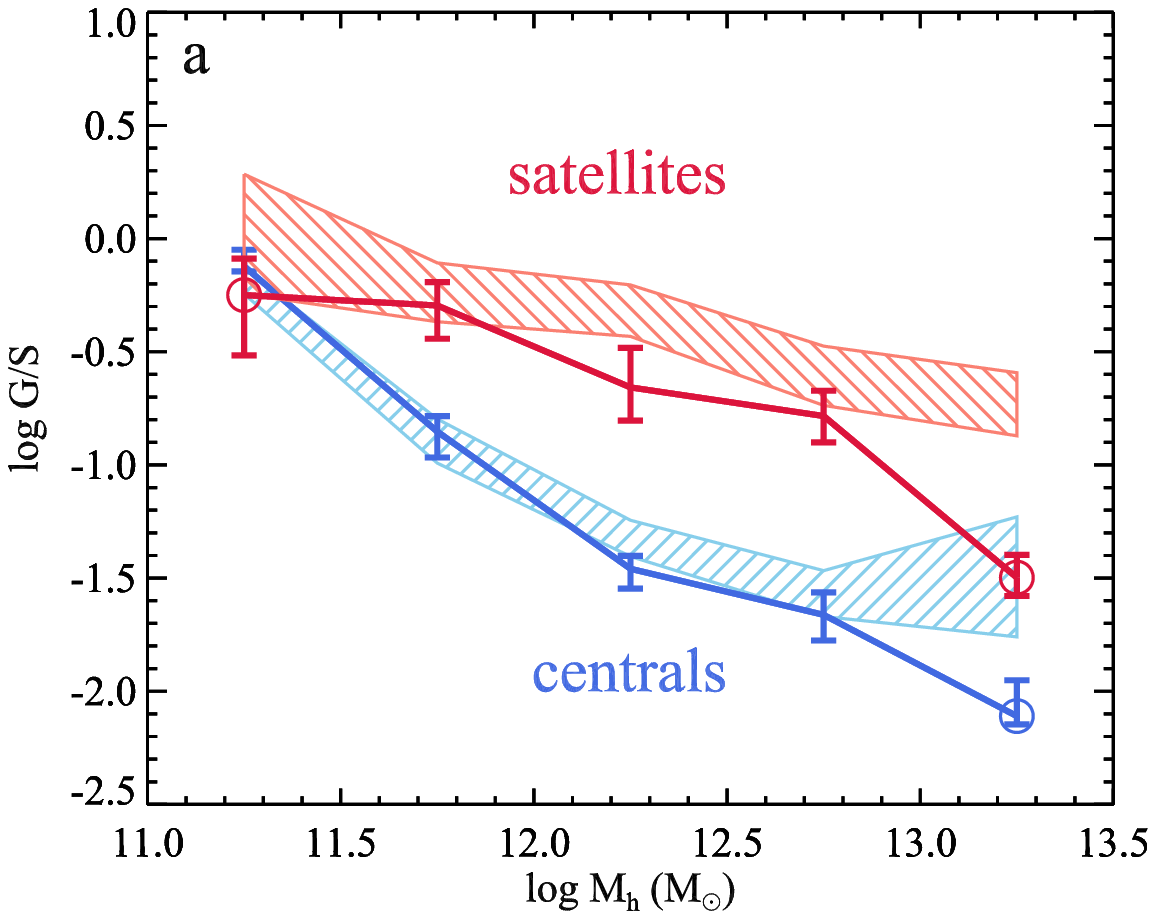}{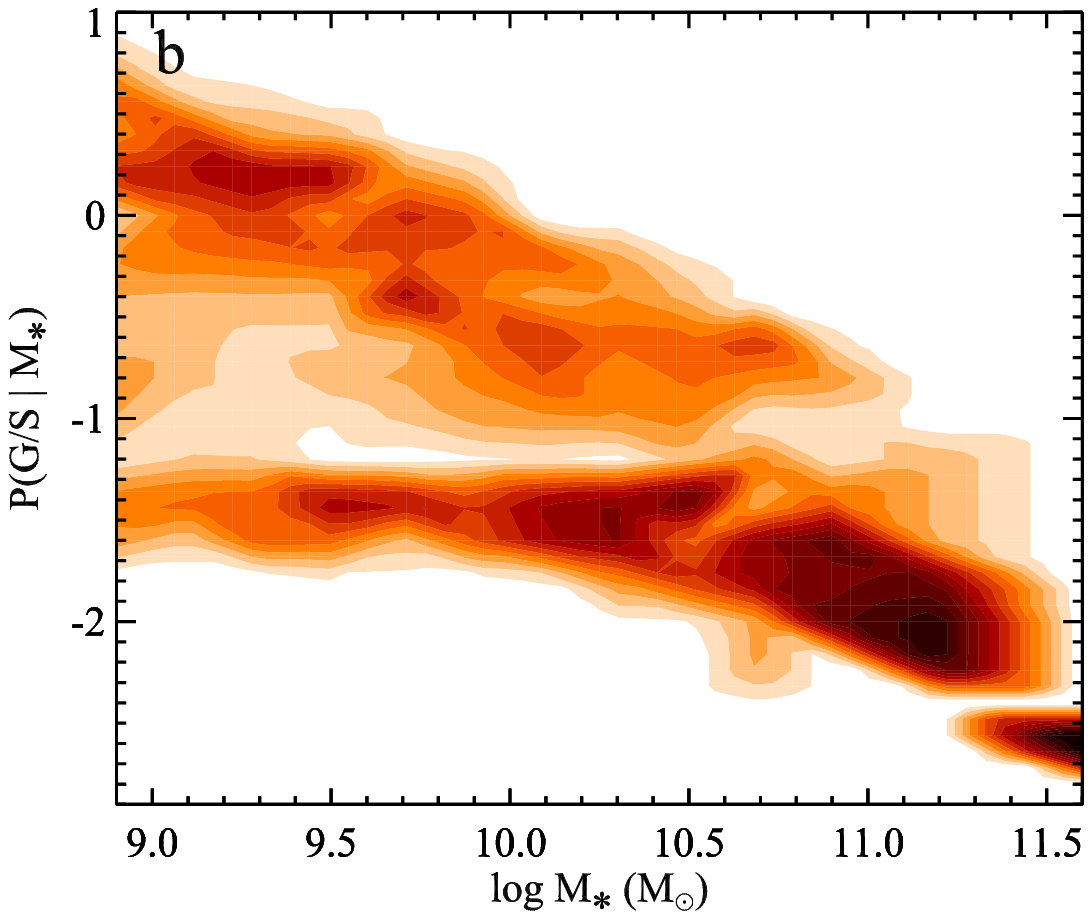}
\caption{\textcolor{black}{(a) Median $G/S$ versus $M_h$ for centrals and satellites.  Medians for satellites consider the {\it integrated} $G/S$ of all satellites in each group within a given bin. Hashed regions show the 68\% confidence bounds on the expected trends based on predicting gas fraction using the distribution of $G/S$ as a function of stellar mass (panel b).  Above $M_h\sim10^{12}\,\msun$ both centrals and satellites have systematically lower $G/S$ than is expected from the stellar mass distribution alone. (b) Contours showing the conditional probability distribution of $G/S$ as a function of $M_*$, $P(G/S \, \vert\ M_*)$, calculated in $M_*$ and $G/S$ bins of 0.2 dex, except above $M_*=10^{11}\,\msun$ where we bin $M_*$ by 0.4 dex.  $P(G/S\, \vert\ M_*)$ is normalized so that the total probability distribution in each $M_*$ bin equals 1. The apparent excess at $G/S \sim 0.05$ is artificial, being the result of the predictor from \citet{Eckert15} which assumes gas-poor galaxies have a tight $G/S$ distribution centered at $0.05$.  }}
\label{fig:ch4:msmhalo_gs_cent_sat_singlepanel}
\end{figure*}

\textcolor{black}{As an alternative way to view the relationship between gas fraction and halo mass, Fig.~\ref{fig:ch4:msmhalo_gs_cent_sat_singlepanel}a shows the median $G/S$ vs.\ $M_h$ relation for centrals and satellites.  The line for satellites does not show the median $G/S$ for all individual satellites in each group halo mass bin.  Instead, we quantify satellite $G/S$ by adding the \HI{} and stellar masses of all satellites in a group and combining them into a total $G/S$ measurement for that group, then take the median of these integrated values in each bin.  }

\textcolor{black}{Hashed regions in Fig.~\ref{fig:ch4:msmhalo_gs_cent_sat_singlepanel}a illustrate the 68\% confidence interval on the expected $G/S$ vs.\ $M_h$ relationship if $G/S$ is predicted using the $G/S$ vs.\ $M_*$ relationship and the distribution of $M_*$ in each $M_h$ bin. This indirect estimation of $G/S$ allows us to understand how the $G/S$ vs.\ $M_h$ relation should behave if there is no environmental influence on $G/S$ whatsoever. To make this prediction, we replace each galaxy's $G/S$ measurement with a value from the probability distribution of $G/S$ as a function of stellar mass, $P(G/S\, \vert\, M_*)$ (Fig.~\ref{fig:ch4:msmhalo_gs_cent_sat_singlepanel}b).   We determine $P(G/S \, \vert\, M_*)$ empirically from the full stellar mass-limited sample, where for each galaxy with stellar mass $M_*$, we estimate the local $P(G/S \, \vert\, M_*)$ using all galaxies with stellar mass within $M_* \pm \Delta M_*/2$ and limited to $M_h>10^{11}\,\msun$.  We set $\Delta M_*=0.2$ except where there are fewer than 20 galaxies in that range, in which case we increase $\Delta M_*$ to 0.4. This increase is only necessary for $M_*\gtrsim10^{11}\,\msun$. However, above $M_*\sim10^{11.2}\,\msun$ there are fewer than 20 galaxies available even with the larger $\Delta M_*$, so our estimate of $P(G/S\, \vert\ M_*)$ may be unreliable (this only affects eight galaxies with $M_h\gtrsim10^{13}\,\msun$).  Each galaxy's $G/S$ measurement is then replaced by a value randomly drawn from $P(G/S \, \vert\, M_*)$, after which we recalculate the median $G/S$ as a function of group halo mass for centrals and satellites. This calculation is repeated 10,000 times.}

\textcolor{black}{In Fig.~\ref{fig:ch4:msmhalo_gs_cent_sat_singlepanel}a, the observed relationship for satellites in halos above $M_h\sim10^{12}\,\msun$ tends to fall slightly below the expected trend based on stellar mass alone (although individual bins do not always show a statistically significant offset on their own, the mean offset averaged over all bins above $M_h=10^{12}\,\msun$ is significant at ${>}3\sigma$). This finding is consistent with satellites in $M_h>10^{12}\,\msun$ groups having lower gas fractions than the general galaxy population at the same stellar mass.  Above $M_h\sim10^{12}\,\msun$ centrals also show a hint (${\sim}2.5\sigma$ significance) of systematically lower $G/S$ than expected based on their stellar mass distribution alone.} 

\subsection{Large-scale Structure}
\label{sec:ch4:lss_gs}

\textcolor{black}{Dark matter halos of the same mass can be found in regions of large-scale structure with widely varying properties (e.g., see Fig.~\ref{fig:ch4:mhalo_dmdens_contour}). In this section, we investigate whether the larger-scale environment around galaxy groups can influence galaxy gas content, or conversely whether gas content is entirely governed by processes on halo scales and below.  We first analyze the link between gas content and large-scale structure density (\S\ref{sec:ch4:dmdens_gs}), which then motivates an analysis of the gas content of low-mass halos in relation to their proximity to significantly more massive groups (\S\ref{sec:ch4:gs_groupdist}).  In \S\ref{sec:ch4:mhalo_gs}, we took care to illustrate how our results can change when different approaches to estimating halo mass are employed.  In our initial analyses of the relationship between gas fraction and large-scale density described below, we proceed using the stellar mass-based halo masses, but in \S\ref{sec:ch4:lss_mhalodef}, we summarize how these results can change for baryonic mass-based halo masses.}

\subsubsection{Large-scale Structure Density}
\label{sec:ch4:dmdens_gs}

\begin{figure*}
\epsscale{1.15}
\plottwo{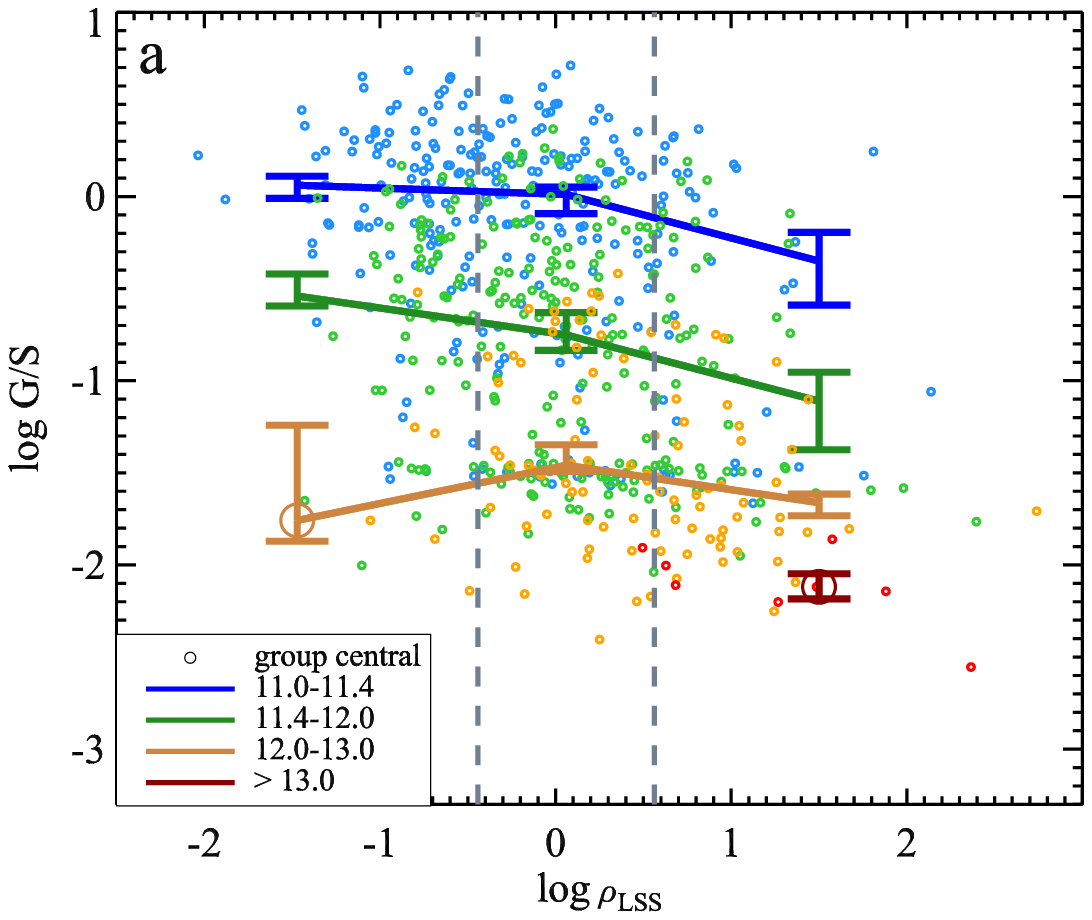}{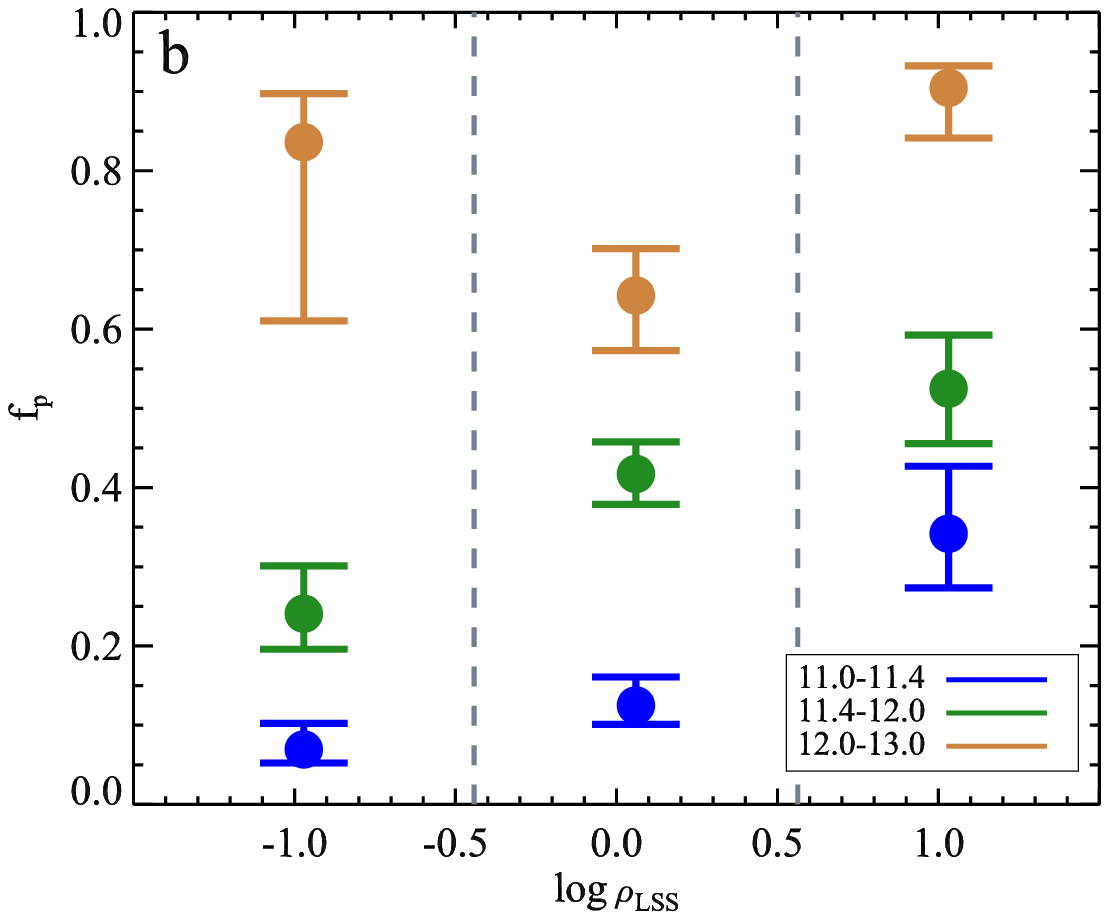}
\caption{\textcolor{black}{(a) $G/S$ versus $\rho_{\rm LSS}$ in fixed group halo mass bins for all group centrals.  Vertical dashed lines delineate  the 25th and 75th percentiles of the $\rho_{\rm LSS}$ distribution.  Points represent individual galaxies and error bars show the 1$\sigma$ confidence intervals on the median value within each $\rho_{\rm LSS}$ regime.  A median is only shown if there are more than 5 points available, and bins with less than 20 points are indicated by large open circles.  In fixed group halo mass regimes, there is a statistically significant correlation between $G/S$ and $\rho_{\rm LSS}$ at $M_h=10^{11-11.4}$ and $10^{11.4-12}\,\msun$. (b) Fraction of gas-poor centrals, $f_p$, in each $\rho_{\rm LSS}$ regime from panel a (where gas-poor is defined as $G/S<0.1$).  Error bars are derived from binomial statistics. We find $f_p$ increases as a function of $\rho_{\rm LSS}$ for $M_h=10^{11-11.4}$ and $10^{11.4-12}\,\msun$}}
\label{fig:ch4:msmhalo_dmdens_gs_all}
\end{figure*}

Fig.~\ref{fig:ch4:msmhalo_dmdens_gs_all}a plots $G/S$ versus $\rho_{\rm LSS}$ for group {\it centrals} (note: only centrals are considered for the majority of this section).  When considering all group halo masses, a Spearman rank correlation test suggests there is a highly significant correlation between $G/S$ and $\rho_{\rm LSS}$.   However, $\rho_{\rm LSS}$ correlates with group halo mass, which in turn correlates with $G/S$.  To remove the influence of group halo mass and isolate the dependence of $G/S$ on only $\rho_{\rm LSS}$, we divide the data into distinct group halo mass regimes ($M_h=10^{11-11.4}\,\msun$, $M_h=10^{11.4-12}\,\msun$, $M_h=10^{12-13}\,\msun$, and $M_h>10^{13}\,\msun$) that are chosen to represent galaxy mass ranges below the gas-richness threshold mass, between the gas-richness threshold mass and the bimodality mass, above the bimodality mass up to what we are calling the large group/cluster scale, and above the large group/cluster scale.  Fig.~\ref{fig:ch4:msmhalo_dmdens_gs_all}a displays the median $G/S$ and its uncertainty within each of these halo mass regimes, further binned into three $\rho_{\rm LSS}$ regimes corresponding to under-dense (bottom 25th percentile of densities), normal-density (middle 50th percentile of densities), and over-dense regions (top 25th percentile of densities). The vertical lines in Fig.~\ref{fig:ch4:msmhalo_dmdens_gs_all}a denote the separations between these $\rho_{\rm LSS}$ regimes. In the $M_h=10^{11-11.4}\,\msun$ and $M_h=10^{11.4-12}\,\msun$ regimes, there are strong correlations (${>}3\sigma$ using a Spearman rank test) between $\rho_{\rm LSS}$ and $G/S$.  The correlation for $M_h=10^{12-13}\,\msun$ is marginal (${\sim}2.5\sigma$). 

\textcolor{black}{As discussed in \S\ref{sec:ch4:dmdens}, $\sim$60\% of groups in the RESOLVE-B sub-volume have densities that require corrections due to edge effects.  To ensure these corrections are not influencing our results, we rerun Spearman Rank correlation tests using only groups that do not require these corrections. With this smaller sample, we still find ${>}3\sigma$ correlations between $G/S$ and $\rho_{\rm LSS}$ in the  \mbox{$M_h<10^{11.4}\,\msun$} and $M_h=10^{11.4-12}\,\msun$ regimes.  However, the statistical significance for \mbox{$M_h=10^{12-13}\,\msun$} falls below $2\sigma$.  Similarly, we test the correlation strengths using just RESOLVE-A, which provides us with a volume-limited data set where only a small percentage of group require corrections for edge effects.  In this case, the statistical significance of the correlation remains ${>}3\sigma$ for \mbox{$M_h<10^{11.4}\,\msun$}, but falls to ${\sim}2.6\sigma$ for \mbox{$M_h=10^{11.4-12}\,\msun$}. For \mbox{$M_h=10^{12-13}\,\msun$} the correlation remains marginal.  We conclude that the link between $G/S$ and $\rho_{\rm LSS}$ is robust for \mbox{$M_h<10^{11.4}\,\msun$}, and not as robust but still likely for \mbox{$M_h=10^{11.4-12}\,\msun$}.  The weaker correlations when using just RESOLVE-A may actually have a physical explanation (see \S\ref{sec:ch4:cosmic_variance} and \S\ref{sec:ch4:cosmic_variance_disc}).}

Inspection of the distribution of points in Fig.~\ref{fig:ch4:msmhalo_dmdens_gs_all}a shows that the correlations between $G/S$ and $\rho_{\rm LSS}$ are largely due to a growing population of gas-poor ($G/S<0.1$) centrals as $\rho_{\rm LSS}$ increases.  To help illustrate this point, Fig.~\ref{fig:ch4:msmhalo_dmdens_gs_all}b plots the fraction of centrals that are gas-poor ($f_p$) broken up into the same group halo mass and $\rho_{\rm LSS}$ regimes as in Fig.~\ref{fig:ch4:msmhalo_dmdens_gs_all}a.  When considering all centrals with $M_h<10^{11.4}\,\msun$ and $M_h=10^{11.4-12}\,\msun$, $f_p$ shows a steady rise with increasing $\rho_{\rm LSS}$.

This discussion has focused entirely on group centrals.  The results of a similar analysis of satellites are less clear as we face much smaller number statistics at the low halo masses where large-scale structure appears to have the largest impact.  For satellites, we find no correlations between $G/S$ and $\rho_{\rm LSS}$ at fixed group halo mass, and $f_p$ is consistent with staying roughly constant at fixed group halo mass. 

\subsubsection{Distance to Nearest $M_h>10^{12}\,\msun$ Group}
\label{sec:ch4:gs_groupdist}

The population of gas-poor centrals at \mbox{$M_h<10^{11.4}\,\msun$} seen in Fig.~\ref{fig:ch4:msmhalo_dmdens_gs_all} is noteworthy because this halo mass regime is expected to have the highest gas accretion rates and to host the most gas-rich galaxies \citep{Keres09,Kannappan13,Nelson13}.  A possible explanation for the existence of these low halo-mass, gas-poor galaxies is that their gas has been stripped by flyby interactions with larger halos, which should lead to gas-poor centrals being found in closer proximity to larger groups compared to gas-rich but otherwise equivalent centrals.  \textcolor{black}{Alternatively, competitive gas accretion or assembly bias correlated with IGM heating could create a similar signature.}

\begin{figure}
\epsscale{1.15}
\plotone{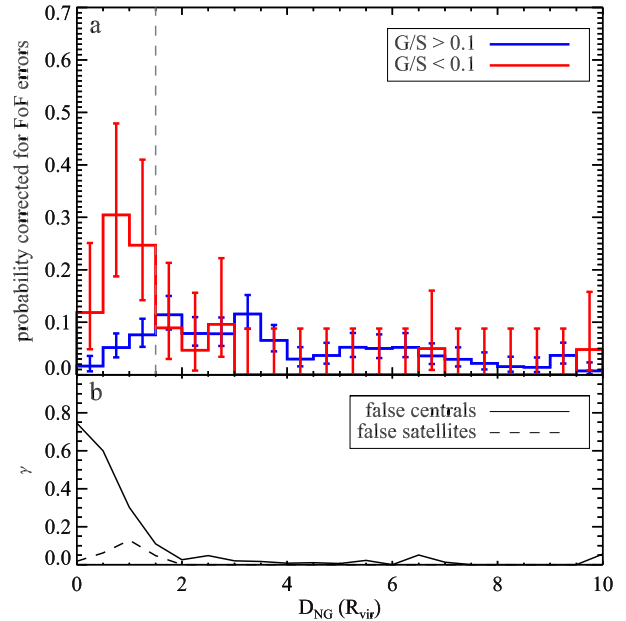}
\caption[Normalized probability distribution of the distances of $M_h<10^{11.4}\,\msun$ centrals from the nearest $M_h>10^{12}\,\msun$ group with recession velocity difference ${<}$500 km s$^{-1}$]{{\bf (a)} Normalized probability distribution of the distances of $M_h<10^{11.4}\,\msun$ centrals from the nearest $M_h>10^{12}\,\msun$ group with recession velocity difference $<$ 500 km s$^{-1}$.  Distances are given as a multiple of $R_{\rm vir}$ of the nearest $M_h>10^{12}\,\msun$ group.  Uncertainties come from Poisson statistics.  The gas-poor population shows a preference to be found within $\sim1.5R_{\rm vir}$ of the nearest $M_h>10^{12}\,\msun$ group. The gray dashed line shows the maximum ``splashback radius" proposed by \citet{More15} as a more physical definition of halo boundaries. {\bf (b)} Errors on central/satellite designation in FoF group finding.  The solid line indicates the fraction of galaxies classified as centrals in $M_h<10^{11.4}\,\msun$ halos that are truly satellites of $M_h>10^{12}\,\msun$ halos.  The dashed line indicates the fraction of satellites in $M_h>10^{12}\,\msun$ halos that are truly centrals in $M_h<10^{11.4}\,\msun$ halos.  These corrections are already applied to panel (a). }
\label{fig:ch4:largermsgroupdist_gs}
\end{figure}

To test these scenarios, in Fig.~\ref{fig:ch4:largermsgroupdist_gs}a we isolate the $M_h<10^{11.4}\,\msun$ population and plot the distribution of their projected distances, $D_{\rm NG}$, from the center of the nearest group with mass ${>}10^{12}\,\msun$.  The specific value of $M_h=10^{12}\,\msun$ was chosen because halos above this scale are more likely to have multiple group members (Fig.~\ref{fig:ch4:groupn_mhalo_hist}) as well as stable hot gas atmospheres \citep{Keres09,Gabor12}, at least one of which may be necessary to strip gas in lower-mass halos\footnote{If we simply examine the distribution of projected distances to the nearest larger group regardless of its specific mass, our results do not change significantly.}. We have already corrected the distributions in Fig.~\ref{fig:ch4:largermsgroupdist_gs}a for the effects of merging and fragmentation by the FoF algorithm (see \S\ref{sec:ch4:group_sep}). These multiplicative correction factors, equal to $1-\gamma$ where $\gamma$ is the false classification rate, are shown in Fig.~\ref{fig:ch4:largermsgroupdist_gs}b.  

 The gas-deficient population is preferentially found within $\sim1.5R_{\rm vir}$ of the closest $M_h>10^{12}\,\msun$ group, where $R_{\rm vir}$ is the virial radius of that massive group's halo\footnote{The mock catalogs used to estimate corrections for fragmentation and blending by the FoF code (\S\ref{sec:ch4:group_sep}) do not reliably predict gas fractions, so we assume the corrections are independent of gas content.  This is likely not correct, since gas-rich and gas-poor galaxies will tend to have different radial distributions in groups (see e.g., \citealt{Geha12}), and the impact of merging and fragmentation on these subpopulations may further vary with large-scale density \citep{Campbell15}.  However, the most conservative test for Fig.~\ref{fig:ch4:largermsgroupdist_gs} is to assume that fragmentation only affects gas-poor galaxies and blending only affects gas-rich satellites.  Under this assumption, we still observe a clear preference for gas-poor centrals to reside closer to nearby $M_h>10^{12}\,\msun$ halos.}. Intriguingly, the radius of $1.5R_{\rm vir}$ within which the gas-poor population is primarily found is equivalent to the maximum  ``splashback radius" discussed by \citet{More15} as a physical definition for the boundary of dark matter halos.  The significance of our results in relationship to the splashback radius is discussed further in \S\ref{sec:ch4:gs_largeenv}.

In Fig.~\ref{fig:ch4:largermsgroupdist_gs}, we ignore any galaxies whose proximity to the edge of the survey volume is smaller than their proximity to the nearest $M_h>10^{12}\,\msun$ group, which removes ~30\% of $M_h<10^{11.4}\,\msun$ centrals from the analysis.  Rejecting these galaxies preferentially removes those that have large values of $D_{\rm NG}$.  To determine whether the gas-rich and gas-poor distributions of $D_{\rm NG}$ are truly distinct even with this bias, we run a Monte Carlo analysis where random values of $D_{\rm NG}$ between the minimum and maximum possible values (see \S\ref{sec:ch4:group_sep}) are assigned to each rejected galaxy. In each Monte Carlo iteration, we calculate two parameters.  First, we run a K-S test to estimate the probability that the distributions of $D_{\rm NG}$ for gas-rich and gas-poor centrals are consistent with coming from the same parent population.  Second, we estimate the value of $D_{\rm NG}$ within which 50\% of gas-rich or gas-poor galaxies are found ($D_{NG,50}$).  Of the 10000 iterations, $>$99.9\% of the time the K-S test says the $G/S<0.1$ and $G/S>0.1$ populations have different distributions of $D_{\rm NG}$ at ${>}3\sigma$. We calculate $D_{NG,50}=1.49\pm0.04$ and $D_{NG,50}=3.00\pm0.07$ for the gas-poor and gas-rich populations, respectively. In summary, centrals in $M_h<10^{11.4}\,\msun$ halos with $G/S<0.1$ are preferentially found closer to their nearest $M_h>10^{12}\,\msun$ group, and this result appears to be robust in the face of both edge effects and possible fragmentation or merging by the FoF algorithm.


\subsubsection{The Impact of Alternative Halo Mass Definitions}
\label{sec:ch4:lss_mhalodef}

Our analysis of the relationship between $G/S$ and $\rho_{\rm LSS}$ has so far been conducted using group dark matter halo masses based on HAM with integrated group stellar mass.  In \S\ref{sec:ch4:mhalo_gs}, we described how the observed relationship between $G/S$, stellar mass, and group halo mass for central galaxies is closely tied to the group parameter used for HAM. We make no assumptions about which parameter is more correct, so it is important to investigate which results are highly dependent on the assumptions built into HAM. To this end, we analyze the relationship between $G/S$ and large-scale environment when estimating halo masses using integrated group baryonic mass instead of integrated group stellar mass.  This analysis again uses the baryonic mass-limited data set with $M_{\rm bary}>10^{9.3}\,\msun$. 

\begin{figure*}
\epsscale{1.15}
\plottwo{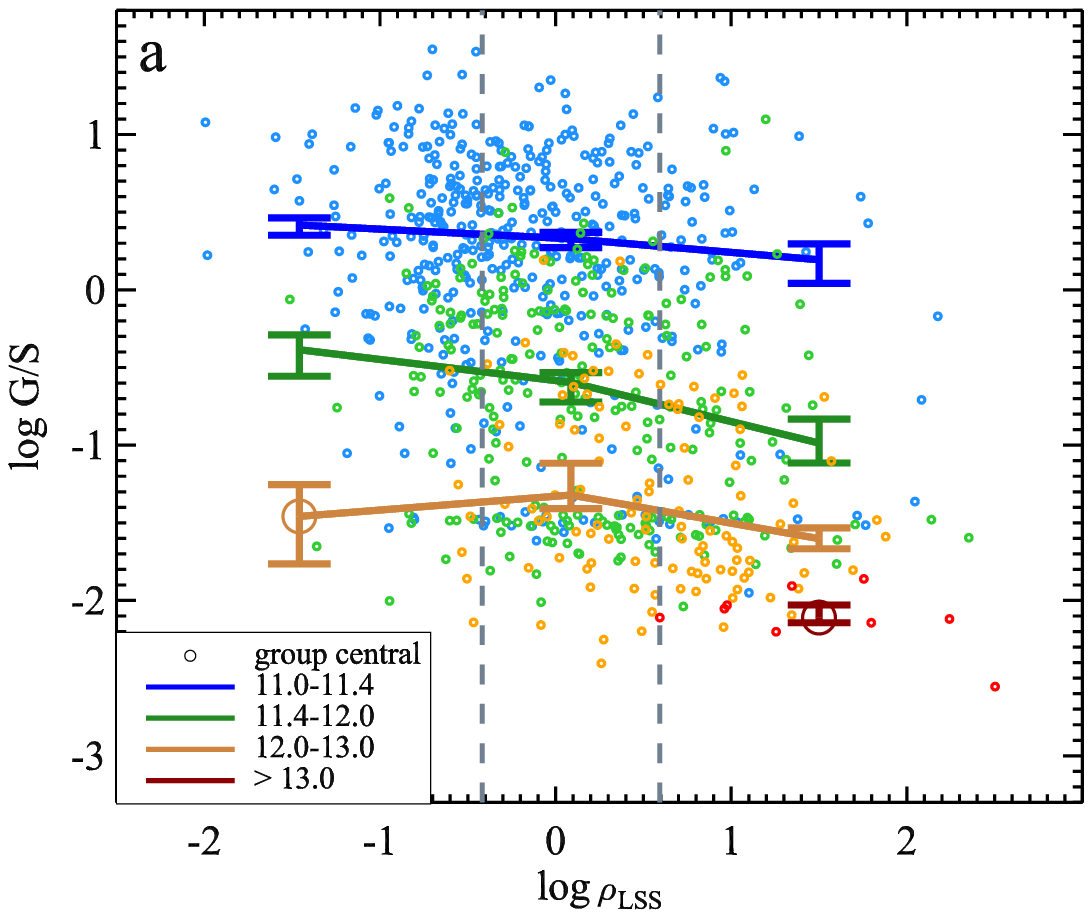}{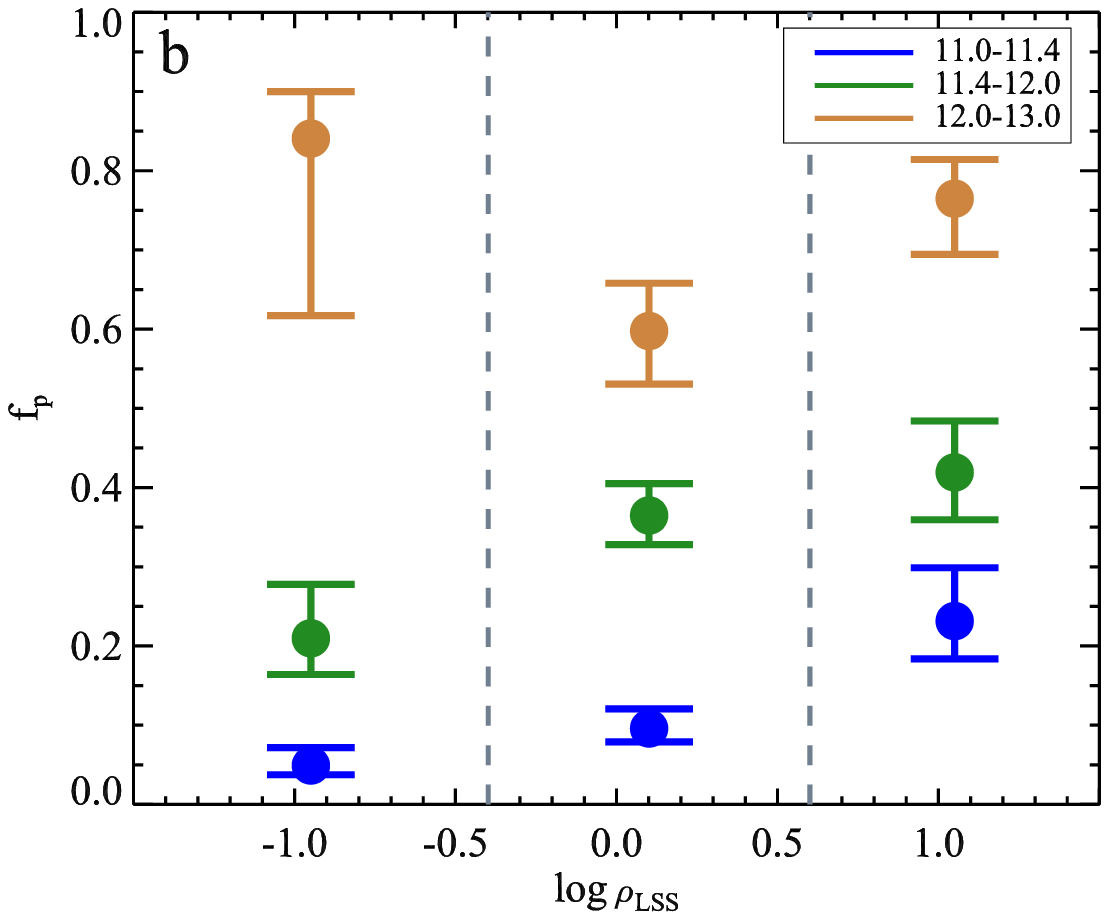}
\caption{\textcolor{black}{Same as Fig.~\ref{fig:ch4:msmhalo_dmdens_gs_all} but using halo masses estimated via group-integrated baryonic mass, rather than stellar mass.  A correlation between $G/S$ and $\rho_{\rm LSS}$ is still present for $M_h<10^{11.4}\,\msun$ and $M_h=10^{11.4-12}\,\msun$ centrals, although at slightly lower statistical significance.  There is still a clear increase in $f_p$ with rising $\rho_{\rm LSS}$ for these same halo mass regimes}.}
\label{fig:ch4:mbmhalo_dmdens_gs_all}
\end{figure*}

\textcolor{black}{As an example to illustrate the effect of using the baryonic mass-limited data set and corresponding halo masses, Fig.~\ref{fig:ch4:mbmhalo_dmdens_gs_all} shows an alternate version of Fig.~\ref{fig:ch4:msmhalo_dmdens_gs_all}, which plots central $G/S$ and $f_p$ as a function of $\rho_{\rm LSS}$ and $M_h$. As in Fig.~\ref{fig:ch4:msmhalo_dmdens_gs_all}a, we find correlations between $G/S$ and $\rho_{\rm LSS}$ for both \mbox{$M_h<10^{11.4}\,\msun$} and \mbox{$M_h=10^{11.4-12}\,\msun$} centrals.  The statistical significances of these correlations are slightly lower than when using the stellar mass-limited sample, but are still above $3\sigma$.  Between \mbox{$M_h=10^{12-13}\,\msun$}, we again find a marginal correlation.  Similarly, Fig.~\ref{fig:ch4:mbmhalo_dmdens_gs_all}b displays a clear increase in $f_p$ with increasing $\rho_{\rm LSS}$.}  

\textcolor{black}{We have re-analyzed our other results from \S\ref{sec:ch4:lss_gs} using baryonic mass-based halo masses, although we do not show them here because the results are very similar to those described above. The behavior of group-integrated satellite $G/S$ as a function of $M_h$ is analogous to that seen in Fig.~\ref{fig:ch4:msmhalo_gs_cent_sat_singlepanel}a where satellites fall systematically below the expected trend predicted from $P(G/S\,\vert\ M_*)$.  However, the mean $G/S$ offset between the measured and predicted trends for centrals above $M_h=10^{12}\,\msun$ is weaker.  Additionally, when using baryonic mass-based halo masses, the difference in $D_{\rm NG}$ distributions for gas-rich and gas-poor $M_h<10^{11.4}\,\msun$ centrals is still present, analogous to Fig.~\ref{fig:ch4:largermsgroupdist_gs}. The Monte Carlo analysis described in \S\ref{sec:ch4:dmdens_gs} suggests that both the $D_{\rm NG}$ distributions and the values of $D_{NG,50}$ for gas-rich and gas-poor centrals are distinct at ${>}3\sigma$ for only 60\% of all iterations, but are distinct at ${>}2.5\sigma$ for ${>}$99.9\% of iterations. $D_{\rm NG}$ values are $\sim30\%$ and $\sim7\%$ larger for gas-poor and gas-rich centrals, respectively.}

In summary, we sometimes find slightly weaker trends between $G/S$ and large-scale environment when using halo masses estimated via baryonic mass, but the statistical significances are not drastically lower and the qualitative results are the same. The weaker trends are likely a side effect of selecting on baryonic mass, which is disadvantageous for studying many of the processes that drive gas depletion. A baryonic-mass selection (and corresponding halo mass estimates based on integrated baryonic mass) leads to more gas-rich and fewer gas-poor galaxies at fixed halo mass. As discussed in \S\ref{sec:ch4:sample}, when examining environmental processes that can lead to lower gas content by gas removal, it is generally more intuitive to compare gas fractions at fixed {\it stellar} mass. However, the analysis with the stellar mass-limited sample may be less appropriate for studying starvation and tidal stripping scenarios.

\subsection{Cosmic Variance}
\label{sec:ch4:cosmic_variance}

\textcolor{black}{RESOLVE is composed of two subvolumes (RESOLVE-A and RESOLVE-B) that span different regions of the local universe with their own large-scale properties.  For example, RESOLVE-B  contains a southern extension of the Perseus-Pisces complex \citep{Giovanelli85b}, it is overabundant in halos with $M_h>10^{13.5}\,\msun$ \citep{Moffett15}, and it has an average galaxy number density of $0.022\,{\rm Mpc^{-3}}$, ${\sim}40$\% larger than RESOLVE-A's number density of $0.016\,{\rm Mpc^{-3}}$ (measured using galaxies with $M_*>10^{8.9}\,\msun$). Given the different properties of the two subvolumes, we explore whether the observed relationships between $G/S$ and environmental properties are consistent between them, and find that there are in fact noticeable dissimilarities.}

\begin{figure}
\epsscale{1.15}
\plotone{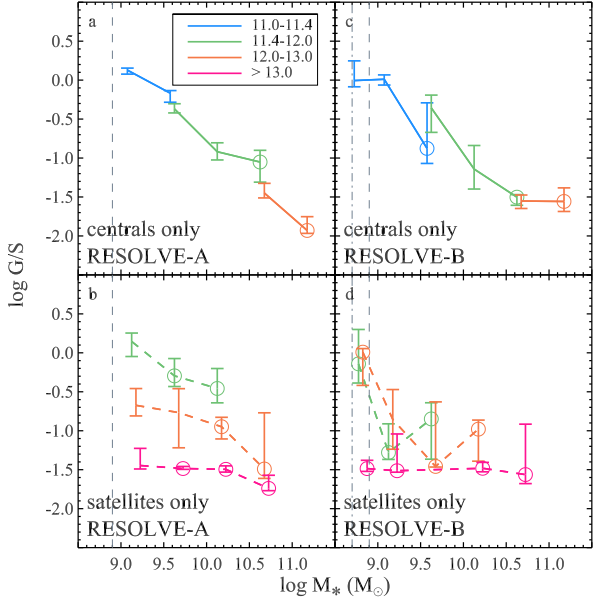}
\caption{\textcolor{black}{Median $G/S$ vs.\ $M_*$ broken up by halo mass regime, central vs.\ satellite, and survey subvolume. The vertical dashed line indicates the completeness limit of $M_*>10^{8.9}\,\msun$ used throughout this work.  The RESOLVE-B subsample has been extended down to its nominal completeness limit of $M_*=10^{8.7}\,\msun$ (denoted by the dashed-dotted line). Points in the same $M_*$ bin are offset slightly for clarity. Satellites in RESOLVE-B above $M_*=10^{8.9}\,\msun$ show no clear dependence on halo mass and appear generally gas poor.  Below $M_*=10^{8.9}\,\msun$, a dependence between satellite $G/S$ and $M_h$ reappears, although only satellites in $M_h>10^{13}\,\msun$ halos appear systematically gas-poor.}}
\label{fig:ch4:medgs_mstars_mbary_sep_springfall}
\end{figure}

\textcolor{black}{Fig.~\ref{fig:ch4:medgs_mstars_mbary_sep_springfall} shows the median $G/S$ vs $M_*$ relation broken up by halo mass for \mbox{RESOLVE-A} and \mbox{RESOLVE-B} separately. For this figure, we have extended the \mbox{RESOLVE-B} subsample down to its true completeness limit of $M_*=10^{8.7}\,\msun$\footnote{\textcolor{black}{Group assignments and halo masses are estimated for this deeper sample following same methodology described in \S\ref{sec:ch4:mhalo}, except we calculate physical linking lengths and the $M_*-M_h$ relation for RESOLVE-B using a version of ECO extending down to $M_*=10^{8.7}\,\msun$}}. Over the same $M_*$ range, the relationships for centrals are consistent between the two subvolumes and RESOLVE-A shows the same trend of decreasing $G/S$ with increasing $M_h$ at fixed $M_*$ reported in \S\ref{sec:ch4:mhalo_gs}, but satellites in \mbox{RESOLVE-B} show no discernible dependence on $M_h$.  Instead, \mbox{RESOLVE-B} satellites appear globally gas poor, even below \mbox{$M_h=10^{12}\,\msun$}, implying group-driven driven gas deficiency may be possible at even lower halo mass scales than discussed in \S\ref{sec:ch4:mhalo_gs}. However, gas-rich satellites are still present at $M_*=10^{8.7-8.9}\,\msun$ in RESOLVE-B, and only those with $M_h>10^{13}\,\msun$ are systematically gas poor.}

\begin{figure}
\epsscale{1.15}
\plotone{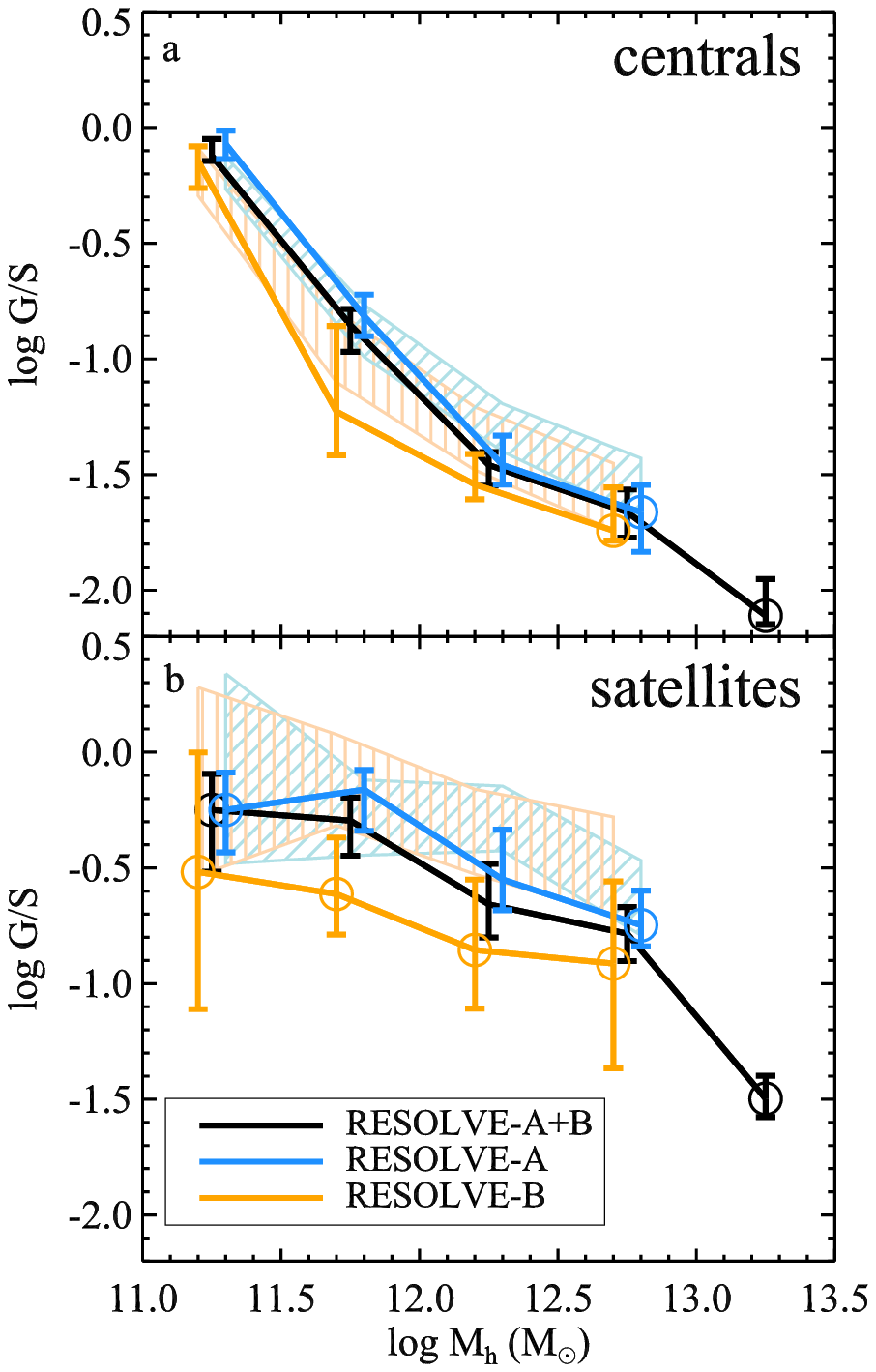}
\caption{\textcolor{black}{Same as Fig.~\ref{fig:ch4:msmhalo_gs_cent_sat_singlepanel}a but showing RESOLVE-A and RESOLVE-B separately. Individual points within the same $M_h$ bin are slightly offset for clarity. The black lines match the combined RESOLVE-A+B data shown in Fig.~\ref{fig:ch4:msmhalo_gs_cent_sat_singlepanel}a. Hashed regions represent the expected trends based solely on the stellar mass distribution in each halo mass bin and the observed $G/S$ vs.\ $M_*$ relationship.  Both centrals and satellites in RESOLVE-B have lower median $G/S$ at fixed halo mass compared to RESOLVE-A.  An offset for centrals is potentially explained by the difference in stellar mass distributions between the two sub-volumes, but the observed difference for satellites is typically larger than can be explained by differences in stellar mass distributions alone.}}
\label{fig:ch4:mshalo_gs_cent_sat_allstruct}
\end{figure}

\textcolor{black}{As an alternative view, Fig.~\ref{fig:ch4:mshalo_gs_cent_sat_allstruct} shows median $G/S$ vs.\ $M_h$ for centrals and satellites in RESOLVE-A and RESOLVE-B separately (as in Fig.~\ref{fig:ch4:msmhalo_gs_cent_sat_singlepanel}, satellite $G/S$ for each group is measured by taking the ratio of the total gas and stellar mass of all satellites in that group). Note that Fig.~\ref{fig:ch4:mshalo_gs_cent_sat_allstruct} does not include the additional \mbox{$M_*=10^{8.7-8.9}\,\msun$} data from RESOLVE-B used in Fig.~\ref{fig:ch4:medgs_mstars_mbary_sep_springfall}. For centrals, the median $G/S$ measured in RESOLVE-B falls below that of RESOLVE-A in all bins, although this difference is only statistically significant at \mbox{$M_h\sim10^{11.5}\,\msun$}. These offsets may be at least partly explained by the different stellar mass distributions in the two subvolumes, as illustrated by the shaded regions in Fig.~\ref{fig:ch4:mshalo_gs_cent_sat_allstruct} (see \S\ref{sec:ch4:mhalo_gs}).  For satellites, we observe a consistent offset that often appears larger than the expected offset from the different stellar mass distributions of satellites in the two subvolumes, although the difference between the RESOLVE-A and RESOLVE-B measurements is technically statistically significant for only $M_h=10^{11.5-12}\,\msun$  (with the additional caveat that uncertainties on the median $G/S$ in RESOLVE-B may not be reliable for due to low number statistics).  Including the RESOLVE-B data down to $M_*=10^{8.7}\,\msun$ slightly increases satellite $G/S$, but the tendency for RESOLVE-B $G/S$ to fall below both RESOLVE-A and the predicted $G/S$ vs.\ $M_h$ relation is still present.}

\begin{figure}
\epsscale{1.2}
\plotone{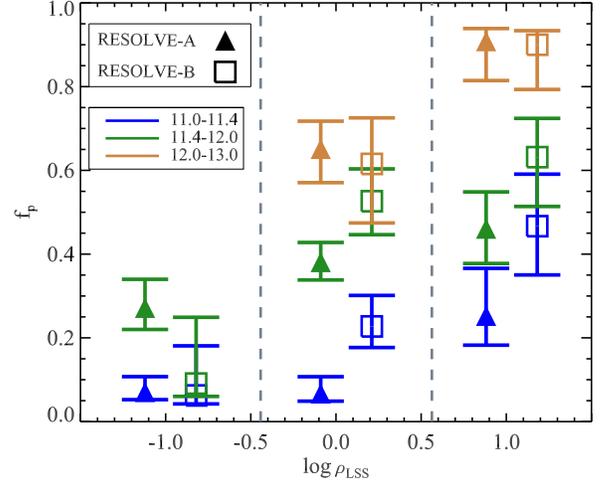}
\caption{\textcolor{black}{Same as Fig.~\ref{fig:ch4:msmhalo_dmdens_gs_all}b except \mbox{RESOLVE-A} and \mbox{RESOLVE-B} data are plotted as separate points. The points are offset slightly for clarity but represent measurements over the same range in $\rho_{\rm LSS}$.  RESOLVE-B shows a stronger relationship between $f_p$ and $\rho_{\rm LSS}$ compared to RESOLVE-A. }}
\label{fig:ch4:quenchfrac_dmdens_msmhalo_sep}
\end{figure}

\textcolor{black}{RESOLVE-A and RESOLVE-B also show differences in the relationship between $G/S$ and $\rho_{\rm LSS}$.  Fig.~\ref{fig:ch4:quenchfrac_dmdens_msmhalo_sep} shows the fraction of gas-poor centrals, $f_p$, as a function of $\rho_{\rm LSS}$ and $M_h$ with RESOLVE-A and RESOLVE-B denoted by different point shapes.  In \mbox{RESOLVE-B}, there is a stronger dependence of $f_p$ on $\rho_{\rm LSS}$ than in \mbox{RESOLVE-A}.  Furthermore, for $M_h<10^{12}\,\msun$ centrals residing in average $\rho_{\rm LSS}$ environments, $f_p$ is larger in RESOLVE-B compared to RESOLVE-A, i.e., the fraction of gas-poor centrals is higher when both $M_h$ {\it and} $\rho_{\rm LSS}$ are fixed.  The behavior of RESOLVE-B does not change significantly if we include galaxies down to its nominal completeness limit of $M_*=10^{8.7}\,\msun$.}

\textcolor{black}{In summary, the relationships between gas content and environmental properties noticeably differ between RESOLVE-A and RESOLVE-B, with RESOLVE-B generally showing a larger fraction of gas-poor galaxies. These results suggest that other properties of the environment, possibly on scales larger than explored in this study, are influencing gas content. We explore this idea further in \S\ref{sec:ch4:cosmic_variance_disc}.}

\textcolor{black}{Alternatively, the different results in RESOLVE-A and RESOLVE-B could arise if RESOLVE-B has a higher rate of incompleteness for gas-rich galaxies.  Given that the ALFALFA survey has been effective at identifying low-luminosity, gas-rich dwarf galaxies missed by other redshift surveys, an incompleteness of gas-rich objects in RESOLVE-B could arise due to the lack of ALFALFA coverage below decl.$\sim 0^{\circ}$. To investigate this possibility, we examine the ratio of galaxies in the baryonic mass-limited ($M_{\rm bary}>10^{9.3}\,\msun$) and stellar mass-limited ($M_*>10^{8.9}\,\msun$) samples, $N_b/N_*$.  For RESOLVE-A and RESOLVE-B, $N_b/N_*$ is 1.20 and 1.02, respectively. We calculate this same ratio in the northern and southern halves of RESOLVE-B (hereafter referred to as RBN and RBS).  If RBS is incomplete in gas-rich galaxies due to the lack of ALFALFA coverage, we would expect $N_b/N_*$ to be significantly smaller for RBS compared to RBN. We calculate $N_b/N_*=$1.06 and 1.03 for RBN and RBS, so RBN has slightly more gas-rich galaxies, but not by a significant amount.  We obtain similar values of $N_b/N_*$ if we extend RESOLVE-B down 0.2 dex to its true stellar and baryonic mass completeness limits.  It is also worth noting that RBS has more high $\rho_{\rm LSS}$ groups than RBN (a K-S test confirms the distributions of $\rho_{\rm LSS}$ are distinct at ${>}3\sigma$ confidence).  Given the observed anti-correlation between $G/S$ in $\rho_{\rm LSS}$, which is observed even if we limit our analysis to just RESOLVE-A, a slightly lower fraction of gas-rich galaxies in RBS compared to RBN is not unexpected.  We conclude that the observed differences between RESOLVE-A and RESOLVE-B are likely real and not the result of preferential incompleteness of gas-rich galaxies in RESOLVE-B.}

\section{Discussion}
\label{sec:ch4:discussion}

Having illustrated the relationship between global galaxy gas fractions and both local and large-scale environment, we now explore the physical processes that may drive these trends.  We first discuss processes associated with dark matter halos, followed by a discussion of physical mechanisms associated with large-scale structure.

\subsection{Drivers of $G/S$ Trends within Halos}
\label{sec:ch4:env_halos}
In \S\ref{sec:ch4:mhalo_gs} (Fig.~\ref{fig:medgs_mstars_mbary_3panel}) we showed how halo abundance matching builds in relationships between stellar mass, $G/S$, and group halo mass for central galaxies.  The resulting bias reduces our ability to discern whether central galaxy $G/S$ decreases smoothly with halo mass, or has more complex behavior.  Such an analysis would require a method of estimating halo masses independently of a group's stellar or baryonic content (e.g., weak lensing). 

Fortunately, we are able to make statements about the satellite population due to behavior that persists independently of the chosen halo mass definition.  Specifically, we show evidence for systematic gas deficiency in satellites residing in halos with masses as low as $10^{12}\,\msun$,  \textcolor{black}{or possibly even lower,} implying that group environmental effects are active well below the large group/cluster scale.  In particular, our results imply the presence of environmentally driven gas deficiency
at group masses \textcolor{black}{at least} one dex lower than the $M_h = 10^{13}\,\msun$
scale probed by \citet{Catinella13}. Our data are also consistent with recent hydrodynamical simulations by \citet{Rafieferantsoa15}, who argue for the emergence of an \HI{}-deficient satellite population starting at $M_h=10^{12}\,\msun$.  Observationally, the onset of lower $G/S$ for satellite galaxies starting at $M_h=10^{12}\,\msun$ was suggested by \citet{Moffett15}, who showed that satellites transition from gas-dominated to star-dominated at approximately this mass scale (their Fig.~23).

\begin{figure*}
\epsscale{1.15}
\plottwo{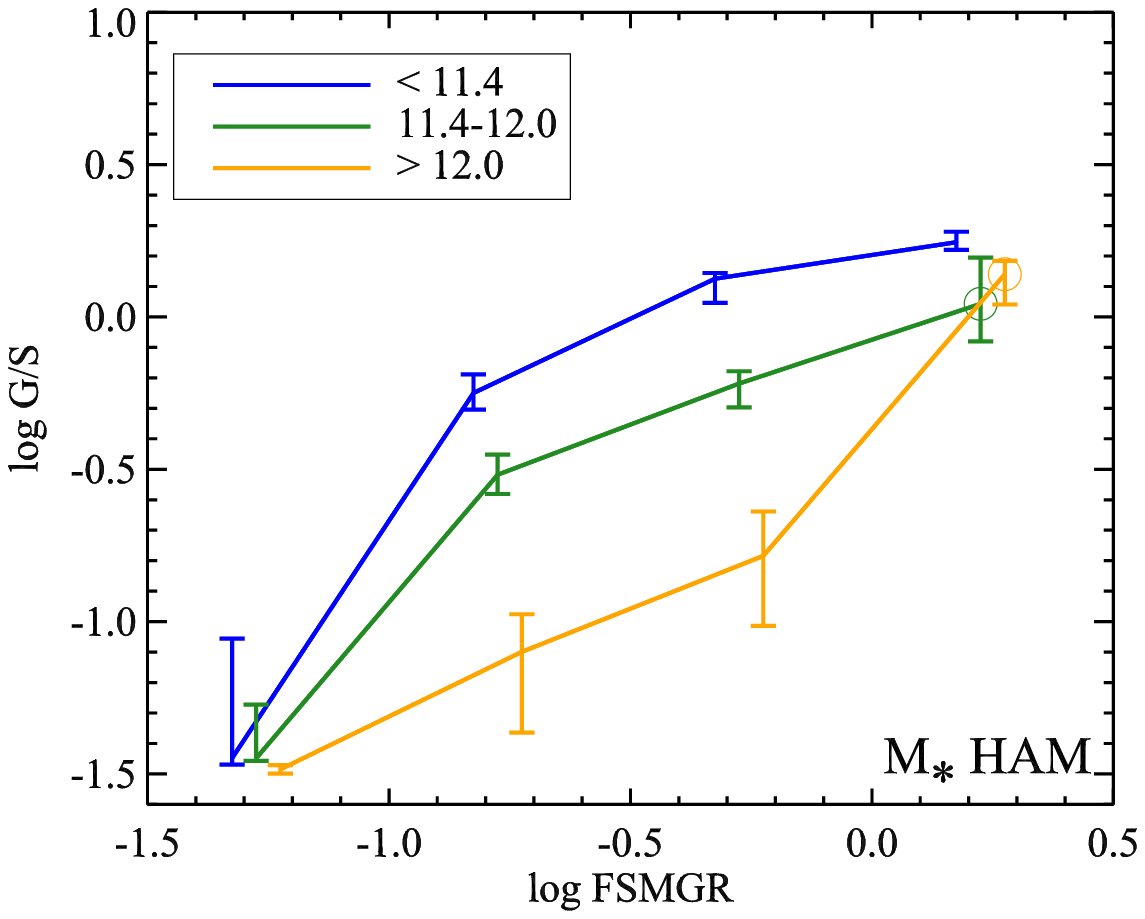}{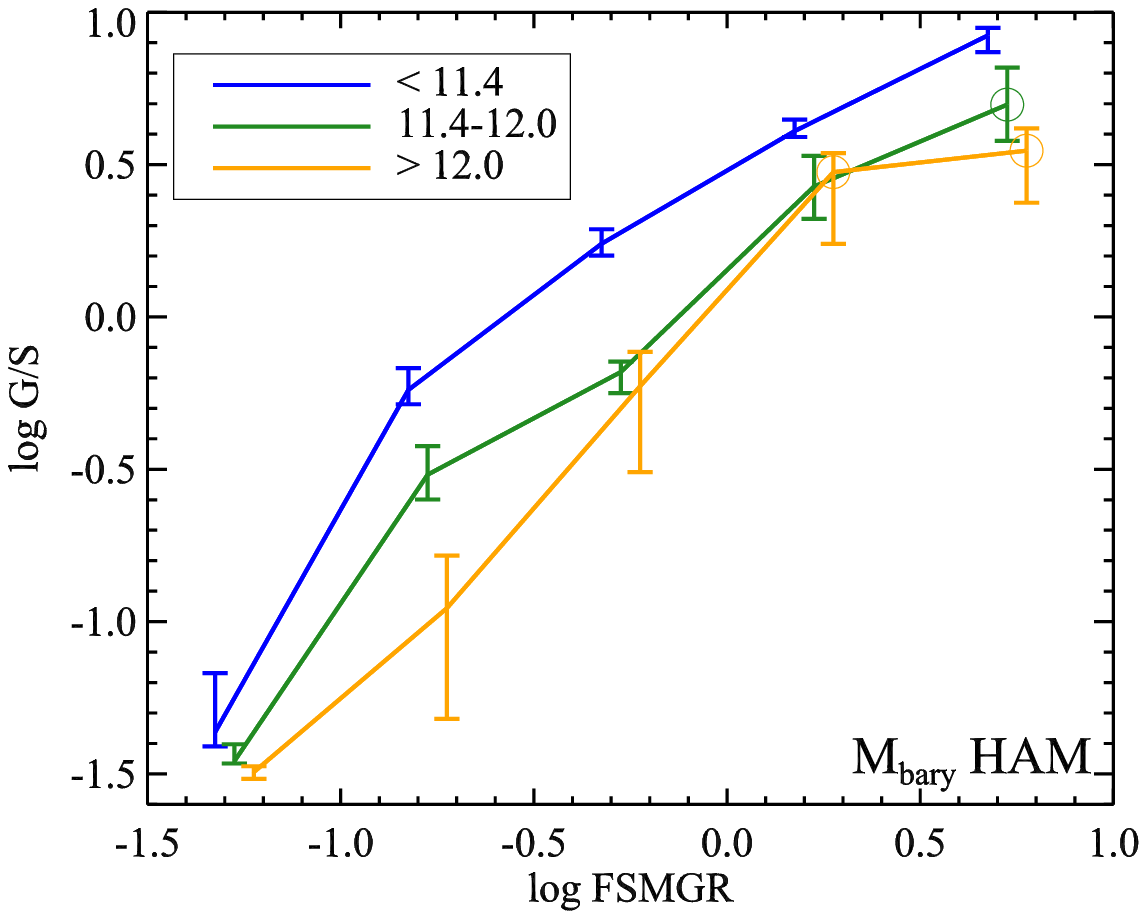}
\caption{Median $G/S$ vs.\ fractional stellar mass growth rate (FSMGR) for galaxies broken up by halo mass (indicated by the legend), for halos estimated from stellar mass (left) and baryonic mass (right).  Points in the same $\log{FSMGR}$ bin are offset slightly for clarify. In both cases, we see evidence that at fixed FSMGR, galaxy $G/S$ decreases as halo mass increases.  This trend could be interpreted as evidence that there is a process directly removing gas from galaxies, or evidence that there is an excess of gas at lower halo masses.}
\label{fig:ch4:medgs_nuvr_mhalobin}
\end{figure*}

Commonly cited physical processes that decrease cold gas content are those that (a) actively remove or consume gas, such as mergers, ram-pressure stripping \citep{Gunn72}, and viscous stripping \citep{Nulsen82}, or \mbox{(b) prevent} the accretion of new gas that would otherwise replenish the gas consumed by star formation (starvation; \citealt{Larson80, Balogh00, Bekki02,Kawata08,Hearin16}).  
 \citet{Catinella13} argue that a ``starvation-only" scenario should cause gas fractions and sSFRs to decline at the same rate, whereas they find that at fixed NUV-$r$ color, gas fractions are systematically lower in more massive halos, implying that a process in addition to star formation is acting directly on the gas reservoir.  Although we do not have a large enough number of galaxies at high halo mass to compare directly to \citet{Catinella13}, we find a similar result at lower halo masses (Fig.~\ref{fig:ch4:medgs_nuvr_mhalobin}), where we have replaced NUV-$r$ color with fractional stellar mass growth rate (${\rm FSMGR}$, see \citealt{Kannappan13}), which is a more direct indicator of star formation history\footnote{The trends in Fig.~\ref{fig:ch4:medgs_nuvr_mhalobin} are similar if we revert to using NUV-$r$ color, although the $M_h=10^{11.4-12}\,\msun$ and the $M_h>10^{12}\,\msun$ regimes are less distinct.}. These results could indicate a process that is acting directly on the gas reservoirs, but they could also be interpreted as evidence for gas excess in lower-mass halos, particularly below $M_h=10^{11.4}\,\msun$ where galaxies may be experiencing overwhelming gas infall rates \citep{Kannappan13}.  To ensure the dependence on halo mass is robust, we have also analyzed the inverse relationship between $G/S$ and ${\rm FSMGR}$ by looking at the effect of halo mass in bins of fixed $G/S$,  finding that a halo mass dependence is only seen for galaxies with $G/S=$0.1 to 1, which may reflect that the most gas-rich systems with $G/S>1$ are not typically found in $N>1$ groups.
 
If a process is depleting gas content faster than star formation alone, there still remains the question of what that process is, be it ram-pressure, viscous, or tidal stripping.   At halo mass scales comparable to $10^{12}\,\msun$, \citet{Kawata08} find that ram-pressure stripping is not efficient enough to remove significant amounts of cold gas.  \citet{Rasmussen08} find similar results from modeling, while also showing no clear connection between \HI{} deficiency and the presence of an X-ray hot gas atmosphere. \citet{Rasmussen08} and \citet{Cluver13} suggest a combination of tidal and viscous stripping may be more relevant for directly removing gas from galaxies in low-mass groups, although these authors focus specifically on Hickson Compact Groups, where such processes driven by interactions may be more pronounced than normal. In broader samples, the role of galaxy mergers and interactions has been questioned.  \citet{Ellison15} argue that mergers do not significantly deplete gas reservoirs and may even lead to gas enhancements (see also \citealt{Rafieferantsoa15}), although the halo mass range of their sample is not reported. \citet{Stark13} find that star formation gradients and ${\rm H_2}$/\HI{} ratios for a broad galaxy sample suggest initial depletion followed by replenishment for blue E/S0 galaxies, but this population is typical of low halo mass environments (${<}10^{12}\,\msun$; \citealt{Moffett15}).   

These considerations lead us to revisit the possible importance of reduced cold gas replenishment, i.e., starvation, above $M_h=10^{12}\,\msun$.  Recently \citet{Wheeler14} found that less than 30\% of $M_*=10^{8.5-9.5}\,\msun$ satellites in $M_h=10^{12.5-14}\,\msun$ groups are quenched, despite predictions from simulations that more than half of them should have been accreted into their host halos more than 6 Gyr ago.  These results imply extremely long gas-exhaustion timescales for satellites in these groups.  Similarly, both \citet{Wetzel15} and \citet{Fillingham15} compare the SFRs of Milky Way satellites to the ELVIS suite of simulations to argue that the quenching timescales of $M_*>10^{9}\,\msun$ satellites in Milky Way-like groups ($M_h\sim10^{12}\,\msun$) are consistent with being driven 
solely by starvation, while direct stripping only becomes important below $M_*\sim10^{8}\,\msun$.  Both \citet{Wetzel15} and \citet{Fillingham15} infer that quenching timescales for satellites are longest at $M_*\sim10^{9}\,\msun$, decreasing at lower stellar mass due to gas stripping, and decreasing at higher stellar mass due to the satellites having had less gas upon entering the system.  In Fig.~\ref{fig:medgs_mstars_mbary_3panel}c, satellites in the lowest stellar mass bin (${\sim}10^{9}\,\msun$) in $M_h=10^{12-13}\,\msun$ halos do not have lower $G/S$ than satellites in $M_h=10^{11.4-12}\,\msun$ halos at a statistically significant level, whereas they do at higher stellar mass. This result may support the starvation picture proposed by \citet{Wetzel15} and \citet{Fillingham15}, where $M_*\sim10^{9}\,\msun$ satellites take the longest to exhaust their gas content.  \textcolor{black}{However, Fig.~\ref{fig:ch4:medgs_mstars_mbary_sep_springfall}b calls this exact interpretation into question given that $M_*\sim10^{9}\,\msun$ satellites in RESOLVE-A have clear gas deficiency in $M_h=10^{12-13}\,\msun$ halos.  Satellites in RESOLVE-B are generally gas poor except at $M_*=10^{8.7-8.9}\,\msun$ (Fig.~\ref{fig:ch4:medgs_mstars_mbary_sep_springfall}d), which may be consistent with the starvation scenario if groups in RESOLVE-B formed earlier than those in RESOLVE-A (see \S\ref{sec:ch4:cosmic_variance_disc} for additional discussion). }

It is noteworthy that we see satellite gas deficiency emerge clearly above the $M_h=10^{12}\,\msun$ scale. This scale is associated with the emergence of stable hot gas atmospheres in dark matter halos \citep{Keres09,Gabor12}, which is an important ingredient for both ram pressure stripping and starvation.  This mass scale is also where the central stellar mass-to-halo mass ratio reaches its maximum and begins to decline \citep{Leauthaud12}, as both satellites and hot gas become more important components of the group.  The observed gas depletion may be related to these changes. \textcolor{black}{However, Fig.~\ref{fig:ch4:medgs_mstars_mbary_sep_springfall} hints that cold gas deficiency may occur in groups below $M_h=10^{12}\,\msun$. K. D. Eckert et al. (in preparation) find a large scatter in halo masses when comparing between dynamical and HAM estimates, particularly around $M_h=10^{11.4-12}\,\msun$, which they term the ``nascent group" regime where groups are forming for the first time.  This scatter in halo mass may be related to varying hot gas fractions, which are also found to be strongly varied in semi-analytic models over the nascent group regime. \citet{Burchett15} find that the detection rate of \ion{C}{4} in the circumgalactic medium declines above $M_h\sim10^{11.5}\,\msun$, consistent with an increase in the amount of hot halo gas above this mass scale. The diversity in $G/S$ observed in our data in this halo mass regime may therefore reflect the inside-out build-up of hot gaseous halos as groups coalesce.}

\subsection{Drivers of $G/S$ Trends within Larger-Scale Environments}
\label{sec:ch4:gs_largeenv}

\textcolor{black}{The results of \S\ref{sec:ch4:lss_gs} show that halo mass alone does not explain the gas content of galaxies.  The fraction of gas-poor centrals, particularly below $M_h=10^{12}\,\msun$, grows steadily with increasing $\rho_{\rm LSS}$ (Fig.~\ref{fig:ch4:msmhalo_dmdens_gs_all}), and low halo-mass gas-poor singleton galaxies are found preferentially close to significantly more massive halos (Fig.~\ref{fig:ch4:largermsgroupdist_gs}).  Our results imply that some mechanism associated with the larger-scale environment around galaxy halos is either directly removing gas, or stopping the re-acquisition of fresh gas. We consider several scenarios to explain our results: flyby interactions, competitive gas accretion, and IGM ram-pressure stripping.}

\textcolor{black}{The preference of gas-poor $M_h<10^{11.4}\,\msun$ centrals to be found close to $M_h>10^{12}\,\msun$ halos supports a scenario in which their low gas fractions are driven by the direct influence of these larger
groups.  Our result is very similar to that of \citet{Wetzel14} who show that the quiescent fraction of galaxies is enhanced within 2.5$R_{\rm vir}$ of $M_h=10^{12-15}\,\msun$ halos (see also results by \citealt{Hansen09}, \citealt{Lu12}, \citealt{Rasmussen12}, \citealt{Wetzel12}).  \citet{Wetzel14} argue that the enhanced  quiescent population is caused by flyby interactions (they refer to flyby galaxies as ``ejected satellites"; other authors have referred to them as ``splashback" or ``backsplash" galaxies, e.g., \citealt{Gill05}), wherein small halos fall into larger ones, are stripped of their gas, and then exit these larger groups at least temporarily.  Simulations by \citet{Sinha12} show that flyby interactions are very common throughout the galaxy population. }

\textcolor{black}{If the flyby interpretation is correct, our results (and those from past studies) show that the influence of a group can extend beyond its virial radius, which raises questions about how to properly define the true extent of dark matter halos. \citet{More15} argue that a physically well-motivated definition for the radius of a dark 
matter halo is the ``splashback radius" (the maximum radius reached by accreting 
matter after its first infall), which more accurately separates material infalling for 
the first time from material that has orbited through the central halo region at least 
once.  The splashback radius of a halo relative to the virial radius is inversely related to the mass accretion rate, and is expected to lie between 0.8 and 1.5$R_{\rm vir}$.  We find that the majority of gas-poor, $M_h{<}10^{11.4}\,\msun$ centrals fall inside $1.5R_{\rm vir}$ of a neighboring $M_h>10^{12}\,\msun$ halo, so one could consider them to already be satellites. }

\textcolor{black}{The majority of potential flyby/splashback galaxies we identify are located in overdense regions. \citet{McBride09} suggest that z=0 growth rates of dark matter halos should be systematically lower in overdense regions compared to underdense regions. Given the link between halo accretion rate and splashback radius, we expect larger splashback radii (relative to $R_{200m}$) and therefore more flyby galaxies in overdense regions, consistent with our results. \citet{Tonnesen15} argue that the central stellar-to-halo mass ratio is higher in groups residing in large-scale overdensities.  Since we use group stellar mass as the default parameter to estimate group halo mass, we may expect halo masses and their virial radii to be overestimated/underestimated in overdense/underdense regions. Therefore, the rate of apparent flybys in overdense regions may actually be larger than we observe (assuming the group extents will be smaller causing some galaxies currently classified as satellites to become centrals), and flybys may actually be distributed even further beyond $R_{\rm vir}$. }

\textcolor{black}{Alternatively, it may be possible for low halo-mass centrals to become gas-deficient without actually passing through more massive groups as in the flyby interaction scenario.  \citet{Hearin16} argue that accretion of material is significantly reduced in dark matter halos with $R_{\rm Hill}<2.5R_{\rm vir}$\footnote{\citet{Hearin16} report a factor of 3, not 2.5.  We give a different coefficient to account for our definition of virial radius, which is 1.2 times larger than the definition of virial radius used by \citet{Hearin16} (A. Hearin, private communication).}, where $R_{\rm Hill}$ is the Hill radius, or the radius within which an object's gravitational field dominates over other bodies.  In such a scenario, a group will consume its gas supply but lose the competition for additional gas to more massive nearby groups.  Following \citet{Hearin16}, we approximate $R_{\rm Hill}$ as
\begin{equation}
R_{\rm Hill}=D\left(\frac{M_{\rm sec}}{3M_{\rm prim}}\right)^{1/3}
\end{equation}
where $M_{\rm sec}$ and $M_{\rm prim}$ are the masses of the secondary halo (the halo for which $R_{\rm Hill}$ is being determined) and the primary halo (a neighboring, more massive halo), and $D$ is the projected separation between the primary and secondary halos.  For every halo, $R_{\rm Hill}$ is estimated using each more massive halo with a recession velocity difference ${<}500\,{\rm km\,s^{-1}}$, and the minimum $R_{\rm Hill}$ from all these neighboring halos is taken as the final value. As discussed in \citet{Hearin16}, this definition is only an approximation of the true Hill radius, but is sufficient to characterize the tidal field of each halo. In our data, we find that the vast majority of gas-deficient systems have $R_{\rm Hill}\lesssim1.5R_{\rm vir}$, although this is a projected value and likely to be larger in 3-dimensional space. Therefore, our gas-poor, low-mass halos reside in environments where we expect them to experience low accretion rates. }

\textcolor{black}{Additional aspects of the larger-scale environment outside halos may contribute to the removal or depletion of cold gas in low-mass groups.  Hydrodynamical simulations by \citet{Benitez-Llambay13} and \citet{Bahe13} show gas loss  in low-mass galaxies as they pass through the filaments and walls of the cosmic web.   This gas loss is attributed to ram-pressure stripping by the IGM.  \citet{Bahe13} show that hot gas, rather than cold gas, is primarily affected by IGM ram-pressure stripping, but the removal of the hot gas deprives galaxies of gas that could otherwise cool and replenish the \HI{} in these systems. Given that the filaments/walls in \citet{Benitez-Llambay13} and \citet{Bahe13} are essentially defined as overdensities,  IGM ram-pressure stripping could contribute to our observed increase in the fraction of gas-poor centrals in overdense regions.  Additionally, \citet{Porter08} show that galaxies moving toward clusters along cosmic web filaments experience enhancements in star formation immediately before falling into clusters, which is likely caused by interactions with other galaxies falling into the same cluster and along the same filament. These star formation enhancements may subsequently contribute to shorter gas depletion timescales and higher fractions of gas-poor galaxies in overdense regions.}

\textcolor{black}{All the mechanisms described above (flyby interactions, competitive accretion, ram-pressure stripping by the IGM) are expected to occur in similar (i.e., overdense) environments and have similar net effects on galaxy gas content, making it difficult to determine whether one or many of these processes are at work.  Comparing our observations to mock catalogs derived from semi-analytic and hydrodynamical simulations in the future may allow use to break this degeneracy.}

\subsection{Explaining the Observed Cosmic Variance}
\label{sec:ch4:cosmic_variance_disc}

\textcolor{black}{Our results from \S\ref{sec:ch4:cosmic_variance} illustrate that the relationship between gas content and environment is not identical in all regions of the $z=0$ universe.  In contrast to RESOLVE-A, satellites in RESOLVE-B show no apparent dependence between their gas fraction and host halo mass, but appear globally gas poor (except below $M_*=10^{8.9}\,\msun$, but there is no RESOLVE-A data in this mass range with which to compare). \mbox{RESOLVE-B} also has a larger fraction of gas-poor centrals, and it displays a stronger correlation between $G/S$ and $\rho_{\rm LSS}$. In \S\ref{sec:ch4:cosmic_variance}, we argued that the observed differences between between our subvolumes are unlikely to be driven by incompleteness in gas-rich or gas-poor galaxies.  Therefore, gas content is being influenced by something we have not yet explored, possibly another environmental property.}

\textcolor{black}{The average environments within RESOLVE-A and RESOLVE-B are not identical. RESOLVE-B is both overabundant in high-mass halos and is overdense relative to RESOLVE-A. RESOLVE-B is also part of the southern extension of the massive Perseus-Pisces complex \citep{Giovanelli85b}.  Given that overdense regions of the universe will be the first to collapse and form structures, and given that RESOLVE-B is overdense relative to RESOLVE-A, one possible explanation for the difference between them is that the latter may on average represent a more evolved region of the universe.  In this sense, the lower gas fractions may be a signature of assembly bias \citep{Gao05,Croton07}, where $G/S$ correlates with halo formation time, analogous to galaxy color/sSFR \citep{Hearin13,Wang13}.  Within this picture, the physical processes driving gas deficiency would not necessarily be fundamentally different in RESOLVE-A and RESOLVE-B, but they would have been active for a longer period of time in RESOLVE-B, leading to an overall higher rate of gas depletion. In addition, earlier forming structures may have experienced more heating of the IGM through gravitational collapse or AGN feedback, increasing cooling times and contributing to lower gas fractions \citep{Cen11,Kauffmann15}.} 

\textcolor{black}{Combining different group formation times with starvation as the primary driver of gas deficiency of satellites can qualitatively explain the different $G/S$ vs.\ $M_*$ relations in our two subvolumes. Gas depletion timescales due to starvation inversely correlate with stellar mass \citep{Fillingham15,Wetzel15}, so the minimum stellar mass of gas-poor satellites in a group will decrease with time (assuming we are in the regime where gas stripping is unimportant).  Satellites in RESOLVE-B are generally gas poor above $M_*=10^{8.9}\,\msun$, whereas RESOLVE-A satellites have higher median gas fractions in this same stellar mass range.  This behavior is consistent with an earlier average group formation time in RESOLVE-B. Comparing our data to semi-analytic models can let us test this explanation in more detail.}

\textcolor{black}{There are additional aspects of the large-scale environment that have not been addressed which could contribute to the observed differences between the two RESOLVE subvolumes.  In particular, we have not examined morphology of the cosmic web (i.e., classification into different geometric features like filaments, walls, nodes, and voids). It is possible that the morphologies of large-scale structures are inherently tied to their relative ages, in that the topological characteristics of large-scale structures are expected to evolve over time \citep{Zeldovich70}. There are  theoretical and observational hints that galaxy assembly is linked to the local morphology of the underlying cosmic web, e.g., the alignment between galaxy angular momentum and large-scale structure \citep{Aragon-Calvo07b, Hahn07, Zhang09b, Codis12, Libeskind12, Trowland13}, or the variation in distributions of SFRs in different regions of large-scale structure \citep{Alpaslan16}.  RESOLVE and ECO, being highly complete and volume-limited, are ideal surveys to examine the influence of large-scale structure geometry and will be the subject of a future publication.}

\section{Conclusions}
\label{sec:ch4:conclusion}

We have presented the first major release of 21cm data for the RESOLVE survey, a multi-wavelength, volume-limited census of galaxies in the local universe complete into the dwarf mass regime and spanning diverse environments.  The census currently provides detections and strong upper limits for $\sim$94\% of RESOLVE. 

We have combined this \HI{} census with metrics designed to characterize galaxy environments on the scales of galaxy groups (dark matter halo mass) and the surrounding cosmic web (large-scale structure density and group separation). We have used this data set to investigate how both halo mass and large-scale environment independently influence the atomic gas content of the $z=0$ galaxy population.  Our key results are as follows:
\begin{itemize}

\item By comparing gas fractions of satellites as a function of stellar mass in different halo mass regimes, we find systematic gas deficiency in groups as low mass as $M_h=10^{12}\,\msun$, \textcolor{black}{and possibly lower} (see \S\ref{sec:ch4:mhalo_gs}, \S\ref{sec:ch4:cosmic_variance},  Fig.~\ref{fig:medgs_mstars_mbary_3panel}, Fig.~\ref{fig:ch4:medgs_mstars_mbary_sep_springfall}).

\item While we find mostly consistent behavior among satellites independent of how we estimate halo mass, an analogous gas fraction--stellar mass--halo mass analysis applied to central galaxies yields results strongly dependent on the integrated group property (stellar or baryonic mass) used in halo abundance matching. We caution that halo abundance matching inevitably builds in relationships for central galaxies (see \S\ref{sec:ch4:mhalo_gs}, Fig.~\ref{fig:medgs_mstars_mbary_3panel}).

\item The fraction of gas-poor ($G/S<0.1$), \mbox{$M_h<10^{12}\,\msun$} centrals grows with increasing large-scale structure density (see \S\ref{sec:ch4:dmdens_gs}, Fig.~\ref{fig:ch4:msmhalo_dmdens_gs_all}).

\item Gas-poor, $M_h<10^{11.4}\,\msun$ centrals at high $\rho_{\rm LSS}$ often reside alone within their halos, but they tend to cluster within ${\sim}1.5R_{\rm vir}$ of the nearest \mbox{$M_h>10^{12}\,\msun$} group.  This result is not driven by fragmentation in group finding, but may indicate a need to revisit the definition of halo boundaries (see \ref{sec:ch4:gs_groupdist}, \S\ref{sec:ch4:gs_largeenv}, Fig.~\ref{fig:ch4:largermsgroupdist_gs}).   

\item Relationships between $G/S$ and large-scale environment are generally independent of whether halo masses are estimated based on stellar or baryonic mass, although the statistical significance of the observed trends is sometimes slightly weaker when using baryonic mass-based halo mass estimates.  We argue that analysis based on stellar mass tends to highlight ram-pressure/viscous stripping interpretations, as opposed to tidal stripping and starvation interpretations, because the latter affects both stellar and gas mass (see \S\ref{sec:ch4:lss_mhalodef}). 

\item \textcolor{black}{The relationship between $G/S$ and environment differ in the two subvolumes of RESOLVE: compared to RESOLVE-A, satellites in RESOLVE-B are more gas poor and the fraction of gas-poor centrals has a stronger correlation with $\rho_{\rm LSS}$. For halos in the middle 50th percentile of densities and with masses $<10^{12}\,\msun$, RESOLVE-B has a larger fraction of $G/S<0.1$ centrals i.e., the gas poor fraction is higher when both $M_h$ {\it and} $\rho_{\rm LSS}$ are fixed. We postulate that this difference may be a signature of assembly bias; RESOLVE-B may be in a more evolved state than RESOLVE-A, and processes that drive gas deficiency have been active for a longer period of time (see \S\ref{sec:ch4:cosmic_variance}, \S\ref{sec:ch4:cosmic_variance_disc}, Fig.~\ref{fig:ch4:medgs_mstars_mbary_sep_springfall}, Fig.~\ref{fig:ch4:mshalo_gs_cent_sat_allstruct}, Fig.~\ref{fig:ch4:quenchfrac_dmdens_msmhalo_sep}).}

\end{itemize}

The results of this work address several of our key questions about the relationship between gas content and environment as presented in \S\ref{sec:ch4:intro}: we find evidence for gas-deficiency of satellites down to $\sim$10$^{12}\,\msun$ halos, which is possibly linked to the emergence of stable hot gas atmospheres in halos at this mass scale.  The hint of gas deficiency down to $M_h=10^{11.4}\,\msun$ may reflect the build-up of hot halo atmospheres from the inner halos outward.  The influence of the group environment may not be limited to galaxies residing within the group itself, as we find evidence that low halo mass (often singleton) galaxies may have their gas content depleted by interactions with more massive halos. Large-scale structure appears to have a substantial influence on gas content, such that large-scale overdensities have higher fractions of gas-poor centrals, which could be attributed to a number of physical processes.

A number of questions remain unanswered, some of which were raised by this study. Can we assess the detailed relationship between gas fraction, stellar mass, and group halo mass for central galaxies without built-in biases from halo mass estimation?  Does satellite gas deficiency begin below $M_h\sim10^{12}\,\msun$?  Can we constrain whether variations in gas content across environment are caused by gas starvation or direct gas removal processes? \textcolor{black}{Does the morphology of large-scale structure play an important role in determining gas fractions? Can we confirm that group/structure formation times explain the different results in our two subvolumes?}

Some of these questions will be the subject of future work with RESOLVE, while others may require future surveys to address. \textcolor{black}{In particular, an analysis of the relationship gas content and large-scale structure morphology will be presented in D. V. Stark et al. (in prep)}. Although we are still working toward a complete physical interpretation of the trends reported in this paper, our results highlight the importance of considering the multi-scale environments of galaxies when developing a complete picture of galaxy assembly.

\section*{Acknowledgments}

We would like to thank the referee, Michael Vogeley, for his careful review which helped improve this work.  We also thank Jessi Cisewski, Manodeep Sinha, Dan Reichart, Gerald Cecil, Fabian Heitsch, Reyco Henning, Alexie Leauthaud, Kevin Bundy, Surhud More, Andrew Hearin, Peter Behroozi, and Duncan Campbell for useful discussions.  We would like to thank the Green Bank and Arecibo scientific staff and
operators for their assistance with our observing program, particularly project friends Tapasi Ghosh, Daniel Perera, Alyson Ford, and Amanda Kepley, as well as Ron Maddalena for his suggestions that improved the efficiency of our GBT observations. The authors acknowledge the work of the entire ALFALFA collaboration team in observing, flagging, and extracting the catalog of galaxies
used in this work.  This work was supported by funding from NSF CAREER grant AST-0955368, the GAANN Fellowship, NC Space Grant Fellowship, the GBT Student Observing Support Program, and the University of North Carolina Royster Society Dissertation Completion Fellowhip. This work was supported by the World Premier International Research Center Initiative (WPI), MEXT, Japan. The National Radio Astronomy Observatory is a facility of the National Science Foundation operated under cooperative agreement by Associated Universities, Inc. The Arecibo Observatory is operated by SRI International under a cooperative agreement with the National Science Foundation (AST-1100968), and in alliance with Ana G. Méndez-Universidad Metropolitana, and the Universities Space Research Association. Funding for SDSS-III has been provided by the Alfred P. Sloan Foundation, the Participating Institutions, the National Science Foundation, and the U.S. Department of Energy Office of Science. The SDSS-III web site is http://www.sdss3.org/. SDSS-III is managed by the Astrophysical Research Consortium for the Participating Institutions of the SDSS-III Collaboration including the University of Arizona, the Brazilian Participation Group, Brookhaven National Laboratory, Carnegie Mellon University, University of Florida, the French Participation Group, the German Participation Group, Harvard University, the Instituto de Astrofisica de Canarias, the Michigan State/Notre Dame/JINA Participation Group, Johns Hopkins University, Lawrence Berkeley National Laboratory, Max Planck Institute for Astrophysics, Max Planck Institute for Extraterrestrial Physics, New Mexico State University, New York University, Ohio State University, Pennsylvania State University, University of Portsmouth, Princeton University, the Spanish Participation Group, University of Tokyo, University of Utah, Vanderbilt University, University of Virginia, University of Washington, and Yale University. This publication makes use of data products from the Two Micron All Sky Survey, which is a joint project of the University of Massachusetts and the Infrared Processing and Analysis Center/California Institute of Technology, funded by the National Aeronautics and Space Administration and the National Science Foundation. This work is based on observations from the UKIDSS survey.  The UKIDSS project is defined in \citet{Lawrence07}.   UKIDSS uses the UKIRT Wide Field Camera (WFCAM; \citealt{Casali07}). The photometric system is described in \citet{Hewett06}, and the calibration is described in Hodgkin et al. (2009). The pipeline processing and science archive are described in Irwin et al (2009, in prep) and Hambly et al (2008). This work is based on observations made with the NASA Galaxy Evolution Explorer. GALEX is operated for NASA by the California Institute of Technology under NASA contract NAS5-98034.



\label{lastpage}

\end{document}